\documentclass[twocolumn,secnumarabic,amssymb, nobibnotes, aps, prb]{revtex4-2}

\usepackage{graphicx}
\usepackage{subfigure}
\usepackage{dcolumn}
\usepackage{bm}
\usepackage{enumitem}
\usepackage{physics}
\usepackage{multirow}
\usepackage{upgreek}

\usepackage{amsmath}
\usepackage{hyperref}
\usepackage{soul}
\usepackage[normalem]{ulem}
\hypersetup{
    colorlinks,
    citecolor=blue,
    filecolor=black,
    linkcolor=blue,
    urlcolor=blue
}
\usepackage{hhline}

\newcommand{\Vg}{$V_\text{g }$}

\makeatletter

\def\maketitle{
\@author@finish
\title@column\titleblock@produce
\suppressfloats[t]}

\makeatother

\begin{document}

\title{Clean quantum point contacts in an InAs quantum well grown on a lattice-mismatched InP substrate}

%


\author{Connie L. Hsueh}
\thanks{These authors contributed equally to this work.}
\author{Praveen Sriram}
\thanks{These authors contributed equally to this work.}
\affiliation{
Department of Applied Physics,Stanford University, Stanford CA 94305, USA}
\affiliation{
Stanford Institute for Materials and Energy Sciences, SLAC National Accelerator Laboratory, Menlo Park, California 94025, USA
}
\affiliation {Geballe Laboratory for Advanced Materials, Stanford University, Stanford, California 94305, USA}
\author{Tiantian Wang}
\affiliation{
Department of Physics and Astronomy, Purdue University, West Lafayette, Indiana 47907, USA}
\affiliation{Birck Nanotechnology Center, West Lafayette, Indiana 47907, USA}
\author{Candice Thomas}
\affiliation{
Department of Physics and Astronomy, Purdue University, West Lafayette, Indiana 47907, USA}
\affiliation{Birck Nanotechnology Center, West Lafayette, Indiana 47907, USA}
\author{Geoffrey Gardner}
\affiliation{Birck Nanotechnology Center, West Lafayette, Indiana 47907, USA}
\affiliation{
Microsoft Quantum Lab Purdue, Purdue University, West Lafayette, Indiana 47907, USA}
\author{Marc A. Kastner}
\affiliation{
Stanford Institute for Materials and Energy Sciences, SLAC National Accelerator Laboratory, Menlo Park, California 94025, USA
}
\affiliation{
Department of Physics, Massachusetts Institute of Technology, Cambridge, Massachusetts 02139, USA}
\affiliation{
Department of Physics,Stanford University, Stanford CA 94305, USA}
\author{Michael J. Manfra}
\affiliation{
Department of Physics and Astronomy, Purdue University, West Lafayette, Indiana 47907, USA}
\affiliation{Birck Nanotechnology Center, West Lafayette, Indiana 47907, USA}
\affiliation{
Microsoft Quantum Lab Purdue, Purdue University, West Lafayette, Indiana 47907, USA}
\affiliation{School of Electrical and Computer Engineering, Purdue University, West Lafayette, Indiana 47907, USA}
\affiliation{School of Materials Engineering, Purdue University, West Lafayette, Indiana 47907, USA}
\author{David Goldhaber-Gordon}
\email{goldhaber-gordon@stanford.edu}
\affiliation{
Stanford Institute for Materials and Energy Sciences, SLAC National Accelerator Laboratory, Menlo Park, California 94025, USA
}
\affiliation{
Department of Physics,Stanford University, Stanford CA 94305, USA}
%
%
\date{\today}

\begin{abstract}
Strong spin-orbit coupling, the resulting large $g$ factor, and small effective mass make InAs an attractive material platform for inducing topological superconductivity. The surface Fermi level pinning in the conduction band enables highly transparent ohmic contact without excessive doping. We investigate electrostatically-defined quantum point contacts (QPCs) in a deep-well InAs two-dimensional electron gas. Despite the 3.3\% lattice mismatch between the InAs quantum well and the InP substrate, we report clean QPCs with up to eight pronounced quantized conductance plateaus at zero magnetic field. Source-drain dc bias spectroscopy reveals a harmonic confinement potential with a nearly $5$ meV subband spacing. We find a many-body exchange interaction enhancement for the out-of-plane $g$ factor $|g_{\perp}^*| = 27 \pm 1$, whereas the in-plane $g$ factor is isotropic $|g^*_{x}| = |g^*_{y}| = 12 \pm 2$, close to the bulk value for InAs.
\end{abstract}
\maketitle

\setcounter{section}{0}

\section{Introduction}


A quantum point contact (QPC) is a ballistic quasi one-dimensional constriction with a tunable conductance, quantized in multiples of $e^2/h$ \cite{van_Houten_1996}. First demonstrated in GaAs/Al$_x$Ga$_{1-x}$As two-dimensional electron gases (2DEGs) over three decades ago \cite{vanWees,Wharam_1988}, QPCs have been incorporated into mesoscale quantum devices for tunnel spectroscopy \cite{Kjaergaard_2016}, quantum dots \cite{Meirav1990}, charge sensors \cite{Elzerman2004, Reilly2007}, electron injectors \cite{vanHouten1988TMF}, spin polarizers \cite{spin-splitter}, electronic beam splitters \cite{Ji2003}, and more. However, demonstrations of clean QPCs in InAs heterostructures remain far fewer.

InAs-based nanostructures have come under a renewed spotlight as a potential platform for proximity-induced topological superconductivity \cite{Deng1557,Fornieri}. InAs has a small effective mass, large spin-orbit coupling, and surface Fermi level pinning \cite{Vurgaftman}. Proximitized by an $s$-wave superconductor and exposed to a magnetic field, a one-dimensional InAs nanostructure should host Majorana zero modes at its ends \cite{Oreg2010, Lutchyn2010, Lutchyn2018}. This makes InAs-based systems an enticing platform for observing and manipulating Majorana zero modes, toward possible eventual topological quantum information processing \cite{Aasen2016,Alicea2011,Hyart2013,Karzig2017}. InAs 2DEGs can be top-down patterned, offering a scaling advantage over directly-grown nanowires for creating complex geometries and for scaling to large numbers of devices \cite{Fornieri, Shabani2016}. The small effective mass $m^* = 0.03m_e$ \cite{ShabaniMIT,SM} in InAs quantum wells (QWs) results in a weak temperature and bias dependence of resistivity, making it easier to decouple the background 2DEG in transport measurements of the QPC. 
Furthermore, a single valley degree of freedom with large bulk $g$ factor $\sim$ 12-15 makes InAs QWs a promising material platform for fast control of spin qubits \cite{MittagDots,Petta_InAs,Stevan_InAs} and quantum simulation of many-body phases \cite{Hensgens2017,Dehollain2020}. Clean QPCs with smoothly tunable transitions are a key building block for integrating InAs quantum dot arrays in quantum simulators and processors. 

We report the investigation of quantized conductance and magnetotransport properties of a narrow gate-defined constriction, fabricated in a buried InAs 2DEG grown by molecular beam epitaxy (MBE) on an InP substrate. The 3.3\% lattice mismatch \cite{Vurgaftman} between InAs and InP leads to a compressive strain on the quantum well and introduces dislocation defects; we demonstrate that despite this, our QPCs are the cleanest amongst the handful of reported works in etched and gate-defined constrictions in InAs and InAs/InGaAs QWs \cite{Debray2009, ShabaniQPC2014, Matsuo2017, Mittag, LeeQPC2019}. The more closely lattice-matched substrate choice of GaSb has been plagued for decades with trivial edge conduction at mesa edges~ \cite{ZurichEdgeState,MicrosoftTrivEdg,Nichele_2016,Thomas2018GaSb} which complicates interpretation of transport measurements. Though purely gate-defined nanostructures have recently allowed circumventing this \cite{MittagMesaGate, Mittag}, InP has superior insulating properties compared to GaSb, simplifying the fabrication and operation of quantum devices. The QPC featured in this paper shows eight pronounced quantized conductance plateaus with a harmonic subband spacing near $5$ meV. The spin-split conductance plateaus in an applied magnetic field let us extract an isotropic in-plane effective $g$ factor $|g_x^*| = |g_y^*| = 12\pm 2$ and an exchange interaction-enhanced out-of-plane $|g_{\perp}^*| = 27 \pm 1$. Our work supports the integration of QPCs into quantum dots and other nanostructures.


\section{Device fabrication and experiment setup}
\label{sec:FabSetup}
The device was fabricated on a heterostructure grown by MBE on a semi-insulating InP (100) substrate; the growth is characterized in detail in Ref.~\onlinecite{Hatke2017} (Sample B). The layer sequence is shown in the cross-sectional schematic Fig.~S1(a) in the Supplemental Material (SM) \cite{SM}. The active region consists of a 4 nm InAs QW sandwiched between 10.5 nm of In$_{0.75}$Ga$_{0.25}$As layers. A 900 nm step-graded buffer of In$_x$Al$_{1-x}$As helps overcome the native lattice mismatch between InP and the quantum well, and a 120 nm In$_{0.75}$Al$_{0.25}$As top barrier moves the active region away from the surface for increased mobility. 

Carriers originating from deep-level donor states in the In$_{0.75}$Al$_{0.25}$As layers populate the 2DEG formed in the InAs QW \cite{Capotondi2004,Luo1993}. The 2DEG has a mobility $\mu = 4.55 \times 10^{5}$ cm$^2$/V$\cdot$s at an electron density $n_s=4.34\times 10^{11}$ cm$^{-2}$ as measured in a 5 $\upmu$m-wide Hall bar at $T = 1.5$ K, corresponding to a mean-free path of $l_{\text{mf}} = 4.9$ $\upmu$m. Owing to suppressed alloy and InGaAs/InAs interface scattering in our buried deep-well heterostructure, the mobility is amongst the highest reported for InAs QWs and is limited by unintentional background impurities and native charged point defects \cite{SM,Hatke2017}. 

Our samples are first processed with standard electron beam lithography and wet etching to define an extended Hall bar-like mesa with an area of 5 $\upmu$m $\times$\, 60\, $\upmu$m between voltage probes. The etch depth is 300 nm, extending into the buffer layer to achieve electrical isolation. To improve surface and edge contact, Ti/Au ohmic contacts are deposited after a light, additional wet etch and \textit{in situ} Ar mill. A 35-nm HfO$_2$ dielectric layer is added by atomic layer deposition at 150 \textdegree C. Finally, Ti/Au gate electrodes are deposited in multiple steps to form pairs of split gates of width 100 nm ($\ll l_{\text{mf}}$) and lithographically designed separations in the range 175 -- 475 nm. The QPC highlighted in this work has separation 325 nm, and data from additional QPCs are included in the SM \cite{SM}. 

\begin{figure}[!htbp] 
            \centering
            \includegraphics[width=0.45\textwidth,keepaspectratio]{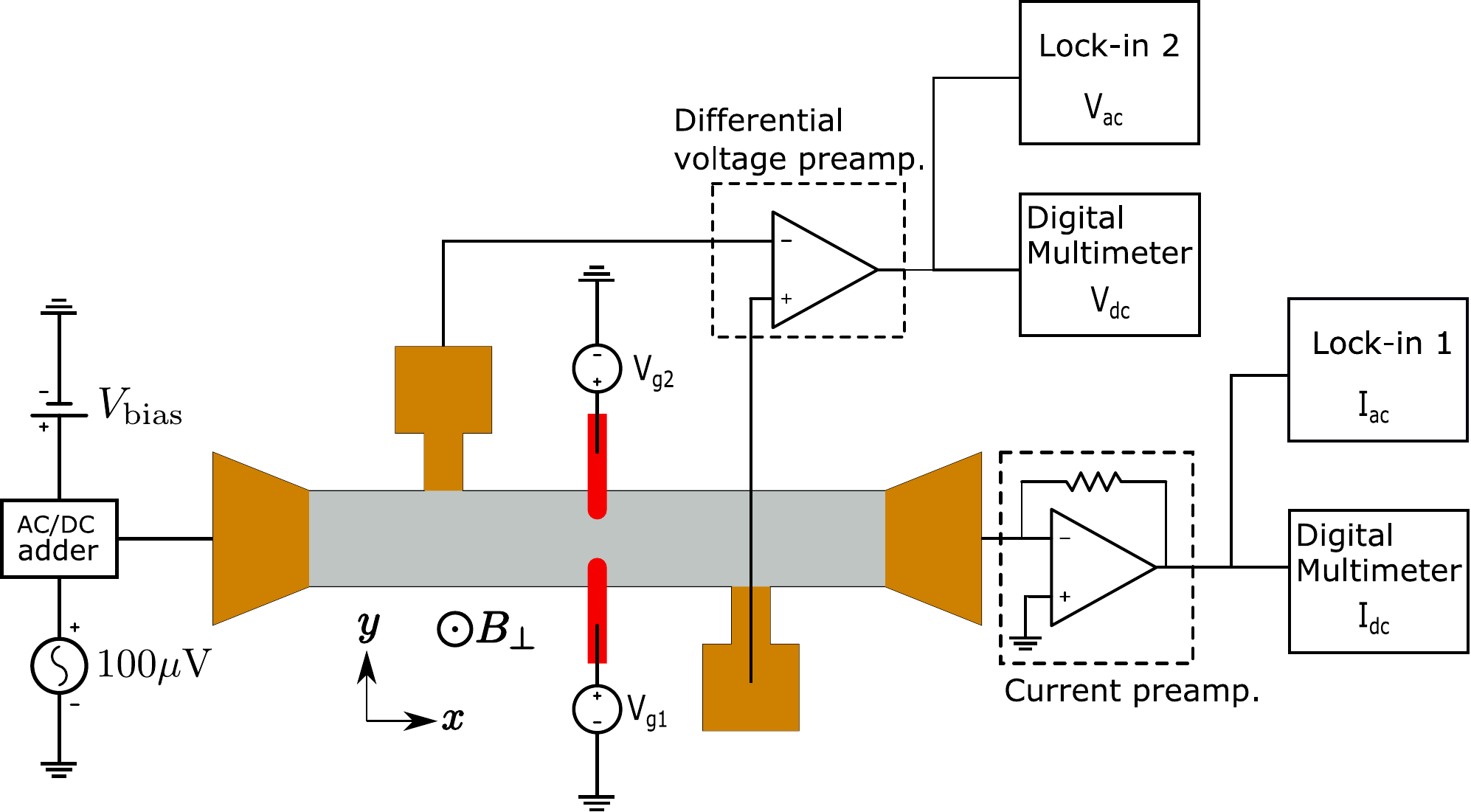}
\caption{Schematic representation of the measurement setup. The quantum point contact (QPC) is defined on a 5 $\upmu$m-wide mesa (gray) with Ti/Au ohmic contacts (yellow). The QPC gates (red) are biased with dc voltage sources $V_{\text{g1,g2}}$. The Hall bar is biased with a 100 $\upmu$V low-frequency ($<20$ Hz) ac  excitation, and a dc bias $V_{\text{bias}}$. The ac  and dc components of the diagonal voltage drop across the QPC are measured after differential amplification by a lock-in amplifier $V_{\text{ac}}$ and digital voltmeter $V_{\text{dc}}$. The voltage probes have a 60 $\upmu$m horizontal separation. A current preamplifier provides a virtual ground at the drain, and a lock-in amplifier and digital ammeter are used to measure the ac  and dc components of the drain current, $I_{\text{ac}}$ and $I_{\text{dc}}$, respectively.}
 	 \label{fig:Schematic}
            \end{figure}

The measurements reported here are performed at $T=1.5$ K in a pumped He-4 cryostat, over multiple cooldowns. A low frequency ($<$20 Hz) ac  excitation of 100 $\upmu$V rms is applied between the source (S) and drain (D) contacts on the extended Hall bar. The current at the drain, $I_{\text{ac}}$, and diagonal voltage drop across the QPC, $V_{\text{ac}}$, are measured using standard low-frequency lock-in techniques. A schematic of the measurement configuration is shown in Fig.~\ref{fig:Schematic}. The conductance through the QPC is $G = \left(V_{\text{ac}}/I_{\text{ac}} - R_{s}\right)^{-1}$, where $R_s$ is the gate-independent 2DEG resistance between the voltage probes.

\section{Results and Discussion}
\label{sec:Results}
\subsection{Conductance quantization}
\label{sec:Gquant}

Negatively biasing the split gates with a voltage around $-1.5 $ V depletes the 2DEG directly underneath, forming a quasi-one-dimensional constriction. Upon further biasing, Fig.~\ref{fig:QPC}(a) clearly shows eight plateaus in $G$ at even multiples of the conductance quantum $e^2/h$ as a function of symmetric gate voltage ($V_{\text{g1}} = V_{\text{g2}} = V_\text{g}$), signifying ballistic transport through the spin-degenerate one-dimensional subbands in the gate-defined constriction, before completely pinching off around $-2.9$ V. Beyond pinch-off, the current is below the noise-floor of the preamplifier $I_{\text{pinch-off}} < 1$ pA, implying a pinch-off resistance $R_{\text{pinch-off}} > 10^8\text{ } \Omega$. The appearance of eight quantized conductance plateaus reveals the pristine nature of the constriction defined by the QPC, and exceeds previous reports \cite{Mittag,LeeQPC2019}. This is compatible with the lithographic split-gate separation $W_{\text{litho}} = 325$ nm and Fermi wavelength $\lambda_F = 36.8$ nm in the 2DEG. The constriction is well described by a saddle-point model in the few-mode limit ($G \leq 8e^2/h)$, as shown in the SM \cite{SM} and the references  \cite{Geier2020,LAUX1988101} therein.
Immediately after cooling the sample, we often observe that pinch-off and other conductance features in $G$ vs $V_{\text{g}}$  gradually drift toward more negative gate voltages.
This could be due to the dynamics of charge traps within the dielectric layer.
After a few days, conductance features in repeated voltage sweeps become reproducible to within a 1 mV relative voltage shift. For consistency, we report data measured with $V_\text{g}$
swept upwards, although once the potential drift stabilizes no significant difference is observed between the two sweep directions. 
\subsection{Finite-bias spectroscopy}
\label{sec:DCbias}
\begin{figure*}[!htbp] 
            \centering
            \includegraphics[width=\textwidth,keepaspectratio]{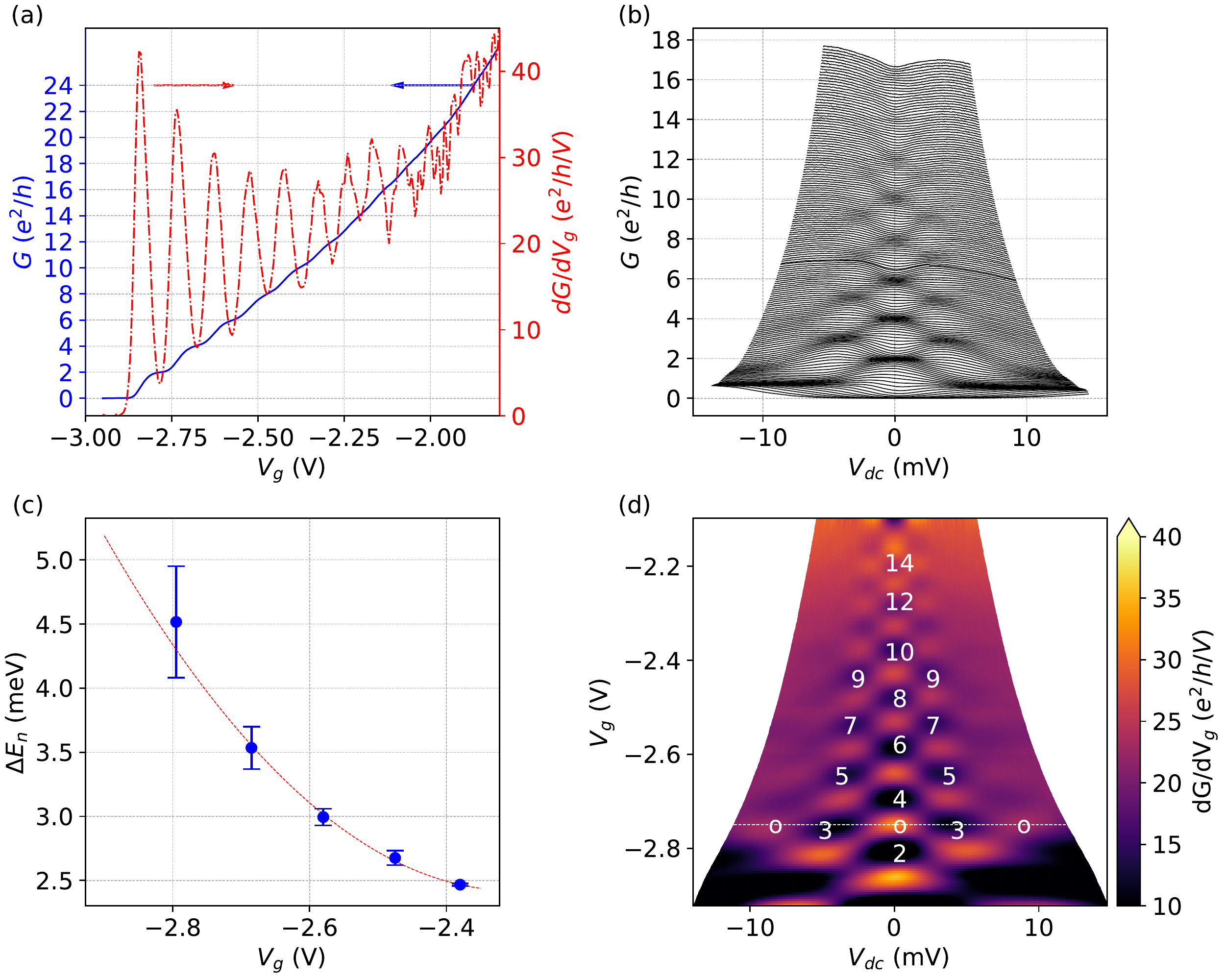}
\caption{(a) Four-terminal conductance $G$ (blue solid line) and transconductance $dG/dV_\text{g}$ (red broken line) through the QPC as the constriction width and local carrier density are modulated by the voltage applied to the split-gates. Quantized conductance plateaus at even-integer multiples of $e^2/h$ are observed. The large number of quantized plateaus visible is an indication of the pristine nature of the QPC. A series resistance $R_s = 380\text{ }\Omega$ has been subtracted to adjust for the 2DEG resistance between the probes. dc bias spectroscopy showing the (b) conductance as a function of $V_{\text{dc}}$, and (d) transconductance as a function of $V_{\text{g}}$ and $V_{\text{dc}}$. Each trace in (b) corresponds to a fixed $V_{\text{g}} \in [-2.92, -2.1]$ V with a step size of 5 mV. A bunching of traces is observed at even multiples of $e^2/h$ around zero bias and at odd multiples at finite bias. The dark regions in (d) correspond to the labeled conductance plateaus in units of $e^2/h$. The bright diamond-shaped stripes of finite transconductance correspond to transitions between the plateaus. A triplet of transconductance maxima, illustrated by the white circles and a dashed horizontal line at \Vg= -2.735 V highlights the harmonicity of the confinement potential. (c) QPC subband spacing plotted as a function of \Vg for the first five subbands. Sweeping the QPC voltages up from pinch-off reduces the curvature of the confinement potential, decreasing the subband spacing. The subband spacings phenomenologically show a quadratic dependence on \Vg.}
 	 \label{fig:QPC}
            \end{figure*}

The level spectrum of the constriction can be probed by applying a dc bias voltage $V_{\text{bias}}$ across the source and drain electrodes of the device. The dc voltage drop across the QPC, $V_{\text{dc}}$, is obtained by subtracting the voltage drop across the bare 2DEG: $V_{\text{dc}} = V_{\text{meas}} - I_{\text{dc}}\times R_s$, where $V_{\text{meas}}$ is the four-terminal dc voltage difference measured across the QPC, $I_{\text{dc}}$ is the dc current through the Hall bar, and $R_s = 380\text{ }\Omega$ is a series resistance arising from the mesa 2DEG resistance. Figure~\ref{fig:QPC}(b) plots $G$ as a function of $V_{\text{dc}}$, where each trace corresponds to a particular $V_{\text{g}}$, as the QPC is opened from pinch-off. A bunching of traces is observed at conductance plateaus, which are even multiples of $e^2/h$ at low-bias, and odd multiples at high-bias. The transconductance $dG/dV_{\text{g}}$ is shown in Fig.~\ref{fig:QPC}(d) as a function of $V_{\text{dc}}$ and $V_{\text{g}}$, with the dark regions corresponding to conductance plateaus and bright regions representing transitions. 

            
The extent of the transconductance diamond for $G=n \times 2e^2/h$ along $V_{\text{dc}}$ is a common measure~ \cite{Rossler_2011} of the energy spacing $\Delta E_n(V_\text{g}^*)$ of QPC subbands $\{n,n+1\}$ at the gate voltage $V_\text{g}^*$ corresponding to the diamond endpoints. Opening the QPC from pinch-off decreases the curvature of the confinement potential, decreasing the subband spacing with \Vg as shown in Fig.~\ref{fig:QPC}(c). The harmonicity of the confinement potential in a particular gate voltage range can be probed by considering a triplet of transconductance maxima circled in Fig.~\ref{fig:QPC}(d). Since they occur at approximately the same gate voltage, we infer $\Delta E_1 \simeq \Delta E_2$ \cite{Rossler_2011}. Similar horizontal lines can be drawn connecting diamond vertices at higher conductances, implying a  harmonic confinement potential, albeit a function of \Vg.

            
Approximating the lateral confinement as a harmonic potential with a gate voltage-dependent angular frequency $\omega_0(V_\text{g})$, the length scale $L_n(V_{\text g}^*)$ of the transverse real-space extent of the subbands at \Vg = \Vg$^*$ can be estimated as
\begin{equation}
    \frac{1}{2}m^*\omega_0^2L_n^2 = \hbar \omega_0\left(n-\frac{1}{2}\right),
\end{equation}
where $m^* = 0.03m_e$ \cite{ShabaniMIT,SM} is the effective mass and $m_e$ is the bare electron mass. Taking $\hbar\omega_0(V_\text{g}^*) = \Delta E_n(V_\text{g}^*)$ for the $n^\text{th}$ subband spacing as determined above, the corresponding length scales can be estimated as $L_1 = 22.7 \pm 0.9$ nm, $L_2 = 45.8 \pm 1.1$ nm and $L_3 = 64.5 \pm 0.8$ nm for the first three subbands, consistent with expectations from the lithographic width $W_{\text{litho}}$ = 325 nm $\gg L_n$.
\subsection{In-plane magnetic field}
\label{sec:Bpar}

\begin{figure*}[!htbp] 
            \centering
            \includegraphics[width=\textwidth,keepaspectratio]{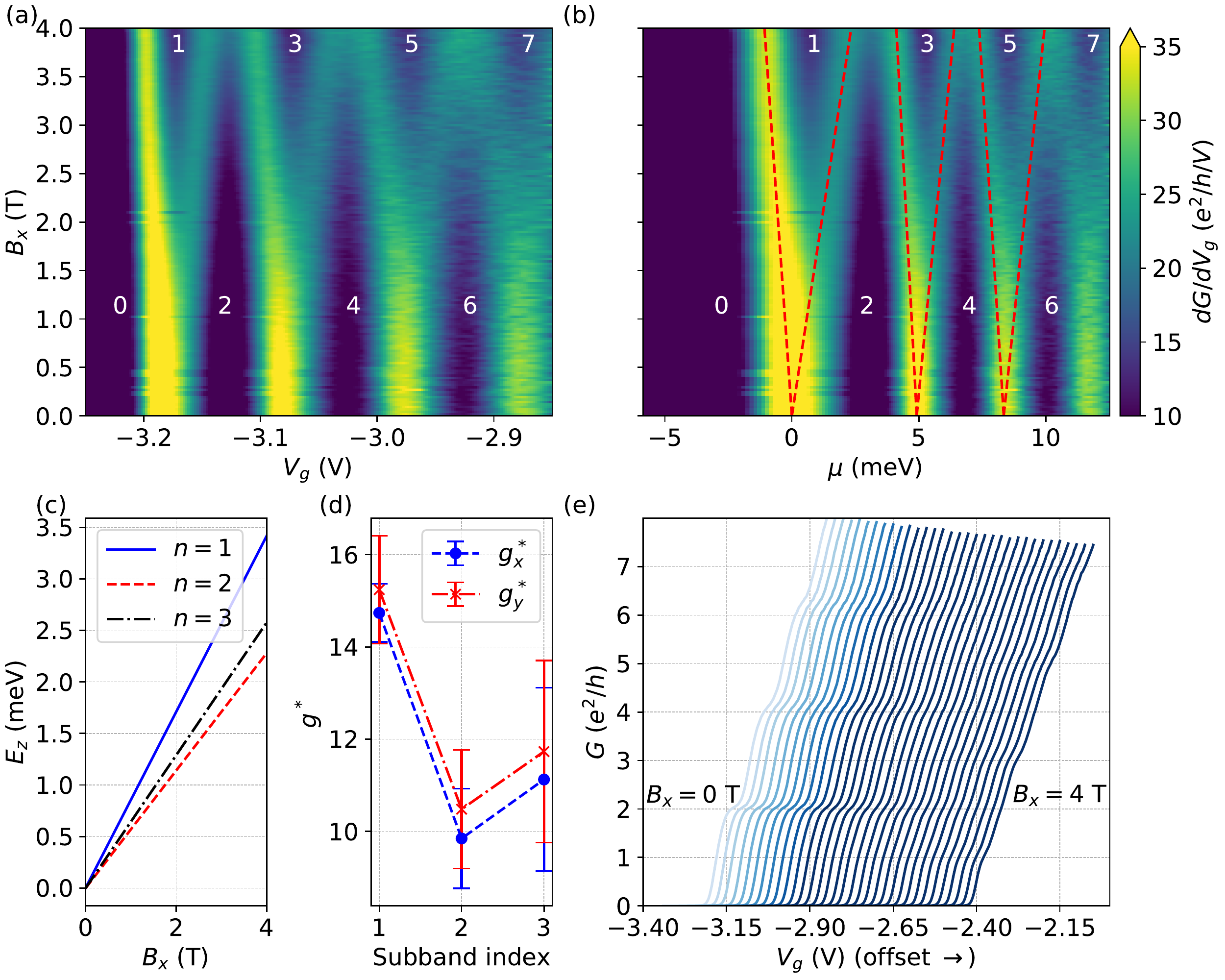}
\caption{
In-plane magnetic field spectroscopy showing (a) transconductance $dG/dV_{\text{g}}$ as a function of QPC gate voltage $V_{\text{g}}$ and a magnetic field $B_{x}$ applied in the plane of the sample and parallel to the transport direction, and (b) as a function of chemical potential $\mu$ estimated from the capacitive lever arm (see the SM \cite{SM}). The dark regions correspond to conductance plateaus labeled in units of $e^2/h$. The transitions between conductance plateaus are visible as bright regions. The appearance of additional dark regions at high $B_{x} (\gtrsim 3 \text{ T})$ is a signature of a Zeeman-induced spin-splitting of the subbands. (c) The Zeeman energy for the first three subbands, extracted from a linear fit to the spin-split transitions (red dotted lines in (b)). (d) The effective in-plane $g$ factor parallel ($g^*_x$) and perpendicular ($g^*_y$) to the transport direction estimated from the slopes of the Zeeman energy in (c) for the first three subbands. Within the error bars, the in-plane effective $g$ factor is isotropic and close to the bulk value for InAs $|g|=13$.
(e) Conductance as a function of $V_{\text{g}}$ at various fixed $B_{x}$. The traces in (e) are offset along the horizontal axis for clarity and display a progressive development of conductance plateaus at $1e^2/h$, $3e^2/h$, and $5e^2/h$.
}
 	 \label{fig:Bpar}
            \end{figure*}
Spin-resolved transport through the QPC can be studied by applying a magnetic field $B_{x}$ in the plane of the sample and parallel to the transport direction.
Figure~\ref{fig:Bpar}(a) shows the transconductance $dG/dV_{\text{g}}$ as a function of $B_{x}$ and $V_{\text{g}}$, as the gate voltage is swept up from pinch-off. The dark, diamond-shaped regions at low $B_{x}$ $(\lesssim 2\text{ T})$ correspond to the spin-degenerate even-integer conductance plateaus. At higher applied $B_{x}$, the spin splitting by the Zeeman effect dominates over the subband linewidths, resulting in the appearance of odd plateaus as additional dark regions interleaved with the spin-degenerate diamonds. Conductance traces as a function of $V_{\text{g}}$ for different $B_{x}$ are shown in Fig.~\ref{fig:Bpar}(e). As expected, conductance plateaus at $1e^2/h$ and $3e^2/h$ emerge as $B_x$ is increased and the width of the even-integer plateaus correspondingly decreases.


Figure~\ref{fig:Bpar}(b) elucidates the spin-split subband spectrum by translating \Vg to a chemical potential $\mu$, using the split-gate lever arm $\alpha = d\mu/dV_{\text{g}}$ extracted from Fig.~\ref{fig:QPC}(d) (see the SM \cite{SM} for details on the conversion). A linear fit to the transconductance maxima for each spin-split subband pair is used to extract the Zeeman energy $E_Z$ as a function of $B_x$, as depicted in Fig.~\ref{fig:Bpar}(c). These linear fits were constrained to intersect at $B_x=0$ for each spin-split subband pair. The in-plane $g$ factor extracted from the slope of the Zeeman energy is shown in Fig.~\ref{fig:Bpar}(d), with error estimates based on fitting parameter variances. 
Figure~\ref{fig:Bpar}(d) also shows the in-plane $g$ factor measured in a magnetic field $B_y$ in-plane but perpendicular to the direction of transport, revealing negligible anisotropy $g^*_{x} \simeq g^*_{y}$ (see the SM \cite{SM}). This is consistent with previous measurements in (In,Ga)As \cite{Martin2010}, InSb \cite{FQu2016} and n-type GaAs \cite{Thomas1996} QPCs. The isotropic in-plane $g$ factor points to a weak Rashba spin-orbit coupling in the constriction \cite{Kolasinki2016}. The absence of intentional dopants and the symmetric In$_{0.75}$Ga$_{0.25}$As barrier structure in the QW stack result in a symmetric 2DEG confinement potential. As revealed by self-consistent Schr\"{o}dinger-Poisson simulations, the QW hosts an electron wavefunction with symmetric tails in the barrier regions [see Fig.~S1(b)] in the SM \cite{SM}). This inversion symmetry of QW in the growth direction [001] leads to the isotropic effective $g$ factor for in-plane magnetic fields{ \cite{Winkler2003}}.
Furthermore, the estimated in-plane $g$ factor $|g^*_{x,y}|$ for the first three subbands $ = \{15 \pm 1, 10 \pm 1, 11 \pm 2\}$ is typical for bulk InAs ($g_\text{InAs} \simeq -13$ \cite{Pidgeon,KONOPKA196729}), with an enhancement for the $n=1$ subband in agreement with theoretical predictions based on exchange interactions \cite{Ando1974,Berggren1996}.

\subsection{Out-of-plane magnetic field}
\label{sec:Bperp}

\begin{figure*}[!htbp] 
            \centering
            \includegraphics[width=\textwidth,keepaspectratio]{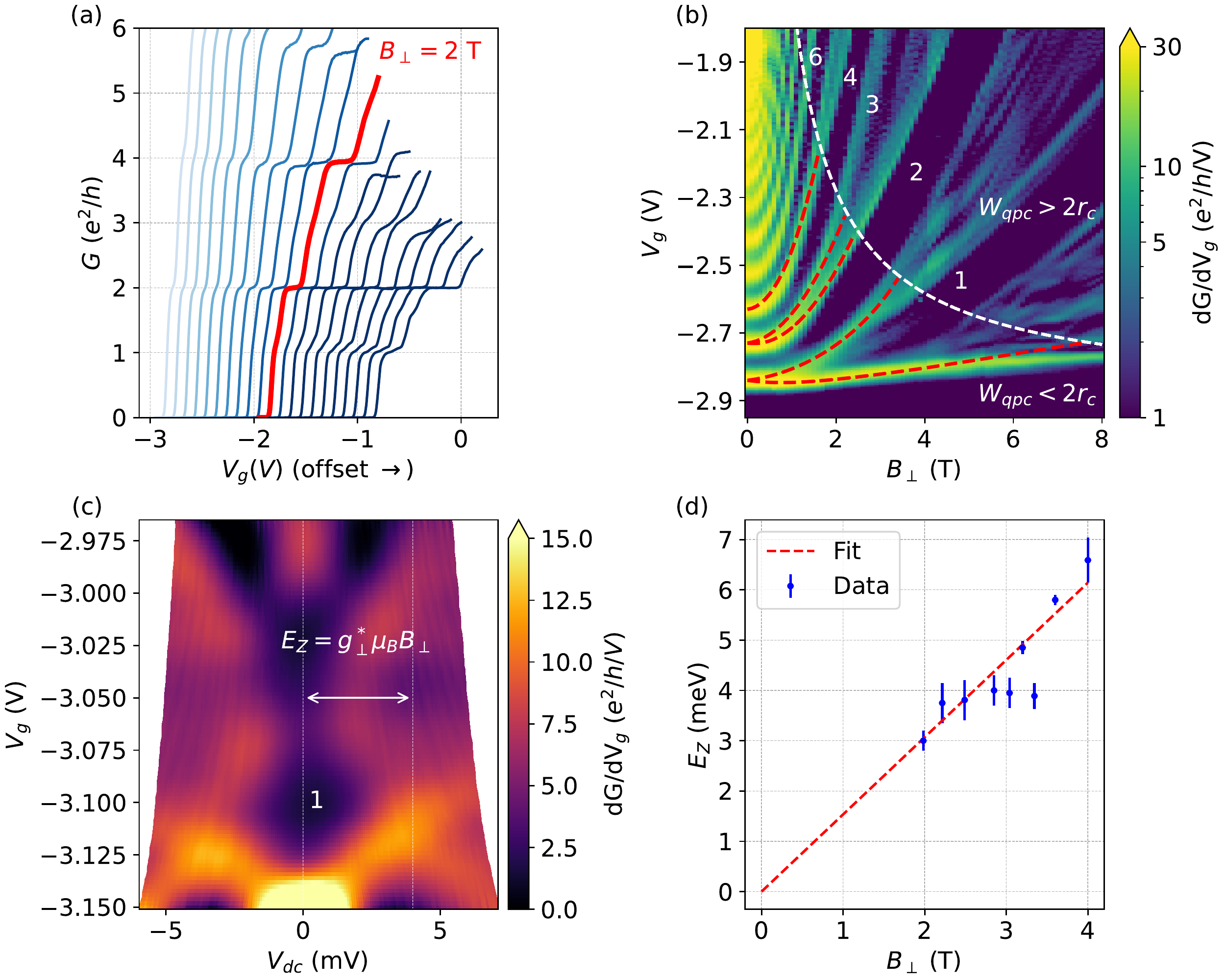}
\caption{(a) Conductance $G$ of the QPC as a function of $V_{\text{g}}$ for a series of out-of-plane magnetic fields. The curves are offset along the horizontal axis for clarity. Above $B_{\perp} = 2$ T (bold red trace), odd-integer conductance plateaus emerge. (b) Transconductance $dG/dV_{\text{g}}$ as a function of QPC gate voltage $V_{\text{g}}$ and a magnetic field $B_{\perp}$ applied out-of-plane of the sample. The dark regions correspond to conductance plateaus labeled in units of $e^2/h$. The transitions between conductance plateaus are visible as bright regions, and illustrate the magnetoelectric subband energy evolution with field. The white dashed curve marks the subband transition to 2DEG Landau levels, based on the gate dependent constriction width $W_{\text{qpc}}(V_{\text{g}})$ and cyclotron radius $r_c$. The red dashed lines show agreement with a model by Beenakker and van Houten \cite{BEENAKKER19911} in the low-field regime $W_{\text{qpc}}<2r_c$. (c) Transconductance as a function of $V_{\text{dc}}$ and $V_{\text{g}}$ at $B_{\perp} = 2.85$ T. The white dashed lines highlight the extent of the $1e^2/h$ conductance plateau diamond along the $V_{\text{dc}}$ axis, a measure of the Zeeman energy $E_Z$. (d) The $B_{\perp}$ dependence of the Zeeman energy, as extracted from the $1e^2/h$ transconductance diamond size, similar to (c). The linear-fit, weighted by inverse $E_Z$ variances, shows an effective out-of-plane $g$ factor $|g^*_{\perp}| \sim  27 \pm 1$.}
 	 \label{fig:Bperp}
            \end{figure*}
            
As a next step in investigating the QPC, we study the effect of electrostatic confinement on magnetic subbands 
by applying a magnetic field $B_{\perp}$ perpendicular to the plane of the sample. The conductance $G$ as $V_{\text{g}}$ is swept up from pinch-off for $B_{\perp}\in [0, 4] \text{ T}$ is shown in Fig.~\ref{fig:Bperp}(a). 
The cyclotron energy of the electrons,  $\hbar\omega_c = \hbar eB_{\perp}/m^*$ where $e$ is the electron charge, adds in quadrature to the QPC confinement energy. The resultant magnetoelectric subbands have a spacing that initially grows quadratically with field $(\omega_y \gg \omega_c)$ before transitioning into a linear increase as they line up with the 2DEG Landau levels for $2r_c < W_{\text{qpc}}$ \cite{BEENAKKER1989127}, where $r_c = \hbar k_F/eB_{\perp}$ is the cyclotron radius of the classical electron trajectory in the 2DEG, $k_F=\sqrt{2\pi n_s}$ is the Fermi wave number in the 2DEG, and $W_{\text{qpc}}(V_{\text{g}})$ is the gate voltage-dependent constriction width (see Sec.~S6 in the SM \cite{SM}). The increase in subband spacing and suppression of backscattering through the Hall bar with $B_{\perp}$  results in broader and more pronounced conductance plateaus. Furthermore, the Zeeman effect of the applied field lifts spin degeneracy and results in the emergence of odd-integer conductance plateaus. Because of thermal ($k_BT \sim 130$ $\upmu e \text{V}$) and disorder broadening in our measurements, we observe spin-split plateaus only at $B_{\perp} \gtrsim 2$ T (red trace).


Figure~\ref{fig:Bperp}(b) depicts the transconductance $dG/dV_{\text{g}}$ as a function of $B_{\perp}$ and $V_{\text{g}}$. The dark regions correspond to conductance plateaus, separated by bright features which represent the transitions between the plateaus. 
The transconductance has a local maximum whenever a subband edge is resonant with the source and/or drain chemical potential.
Given that the confinement is described by a \Vg-dependent harmonic potential, the magnetoelectric subbands can be described by the Beenakker and van Houten model \cite{BEENAKKER19911}

\begin{equation}
    E_{n,\pm} = E_0 + \left(n-1/2\right)\hbar\sqrt{\omega_{y}^2(V_{\text{g}}) + \omega_c^2} \pm \frac{1}{2}g^*_{\perp}\mu_BB_{\perp},
    \label{eq:BvH}
\end{equation}
where $n=1,2,\dots$ is the spin-degenerate subband index, $\pm$ labels the spin-split subband with spin oriented antiparallel (parallel) to $B_{\perp}$, $E_0$ is the energy offset of the conduction band edge, $\hbar \omega_{y}=\Delta E_n$ is the \Vg-dependent QPC subband spacing at $B_{\perp} = 0 \text{ T}$ (see Fig.~\ref{fig:QPC}(d)), and $g^*_{\perp}$ is the effective out-of-plane $g$ factor. The white dashed curve in Fig.~\ref{fig:Bperp}(b) marks the contour  $W_{\text{qpc}}(V_\text{g}) = 2r_c$. An agreement to Eq.~\ref{eq:BvH}, when translated to gate voltage, in the low-field regime $\left(B_{\perp}  < 2\hbar k_F/eW_{\text{qpc}}\right)$ for $n\in \{1,2,3\}$ is shown as red dotted lines in Fig.~\ref{fig:Bperp}(b). The spin-degenerate part of Eq.~\ref{eq:BvH} is used for the $n=3$ subband edge since spin splitting is not well observed for $W_{\text{qpc}}<2r_c$.

The Zeeman energy can be measured by performing finite-bias spectroscopy of the QPC as a function of $B_{\perp}$. In a setup identical to Sec.~\ref{sec:DCbias}, the QPC conductance is measured as a function of an applied dc voltage at a fixed $B_{\perp}$. Figure~\ref{fig:Bperp}(c) shows the transconductance as a function of the dc voltage drop across the QPC $V_{\text{dc}}$ and $V_{\text{g}}$ around the $G=1e^2/h$ plateau at $B_{\perp}=2.85$ T. The dark highlighted region corresponds to the $1e^2/h$ plateau, the extent of which along $V_{\text{dc}}$ corresponds to the Zeeman energy $E_Z = g^*_{\perp}\mu_BB_{\perp} = 4$ meV. Measured as a function of $B_{\perp}$, Fig.~\ref{fig:Bperp}(d) shows the Zeeman energy evolution with field which fits a straight line constrained to pass through the origin, for $|g_{\perp}^*| = 27 \pm 1$. The uncertainty in ascertaining the boundaries of the $G=1e^2/h$ plateau, as evinced by broadened transconductance peaks in Fig.~\ref{fig:Bperp}(c), results in large error bars for the Zeeman energies, defined as the width corresponding to 99\% relative peak height. This can also be seen from the broad transconductance peaks in the $B_{\perp} \in [2,4]$ T region of Fig.~\ref{fig:Bperp}(b). Nevertheless, we can report a two-fold enhancement of the out-of-plane $g$ factor compared to the in-plane and bulk InAs value $g^*_{\perp}/g^*_{x,y} \approx 2$.

The reduced symmetry in quasi-2D heterostructures, as compared to the bulk, introduces anisotropy between $g^*_{\perp}$ and $g^*_{x,y}$ \cite{Winkler2003}. Furthermore, as previously measured \cite{Martin2010} and analyzed \cite{Kolasinki2016} for (In,Ga)As QPCs, the orbital effect of the out-of-plane field strengthens many-body exchange interactions in the 2DEG, resulting in an enhanced $g^*_{\perp}$. The depopulation of consecutive spin-split Landau levels with $B_{\perp}$ leads to an oscillatory exchange enhancement, with local maxima at odd filling factors \cite{Nicholas1988,Sadofyev2002,Cho2004}. \citeauthor{Sadofyev2002} \cite{Sadofyev2002} measured an enhanced out-of-plane $g$ factor $\simeq 60$ at high fields in InAs/AlSb QWs. Similar measurements for $g^*_{\perp}$ in our QW reveal an enhanced 2DEG $g$ factor $\simeq 30$ in the $B_{\perp}\in [2,4]$ T field range (see the SM \cite{SM}). Consequently, we attribute the enhanced splitting of the QPC subband [Fig.~\ref{fig:Bperp}(c) and~\ref{fig:Bperp}(d)] to many-body exchange interactions in the 2DEG, rather than 1D confinement effects due to the constriction.

            
\subsection{Shifting the confinement potential}
\label{sec:Asym}
\begin{figure}[!htbp] 
            \centering
            \includegraphics[width=0.45\textwidth,keepaspectratio]{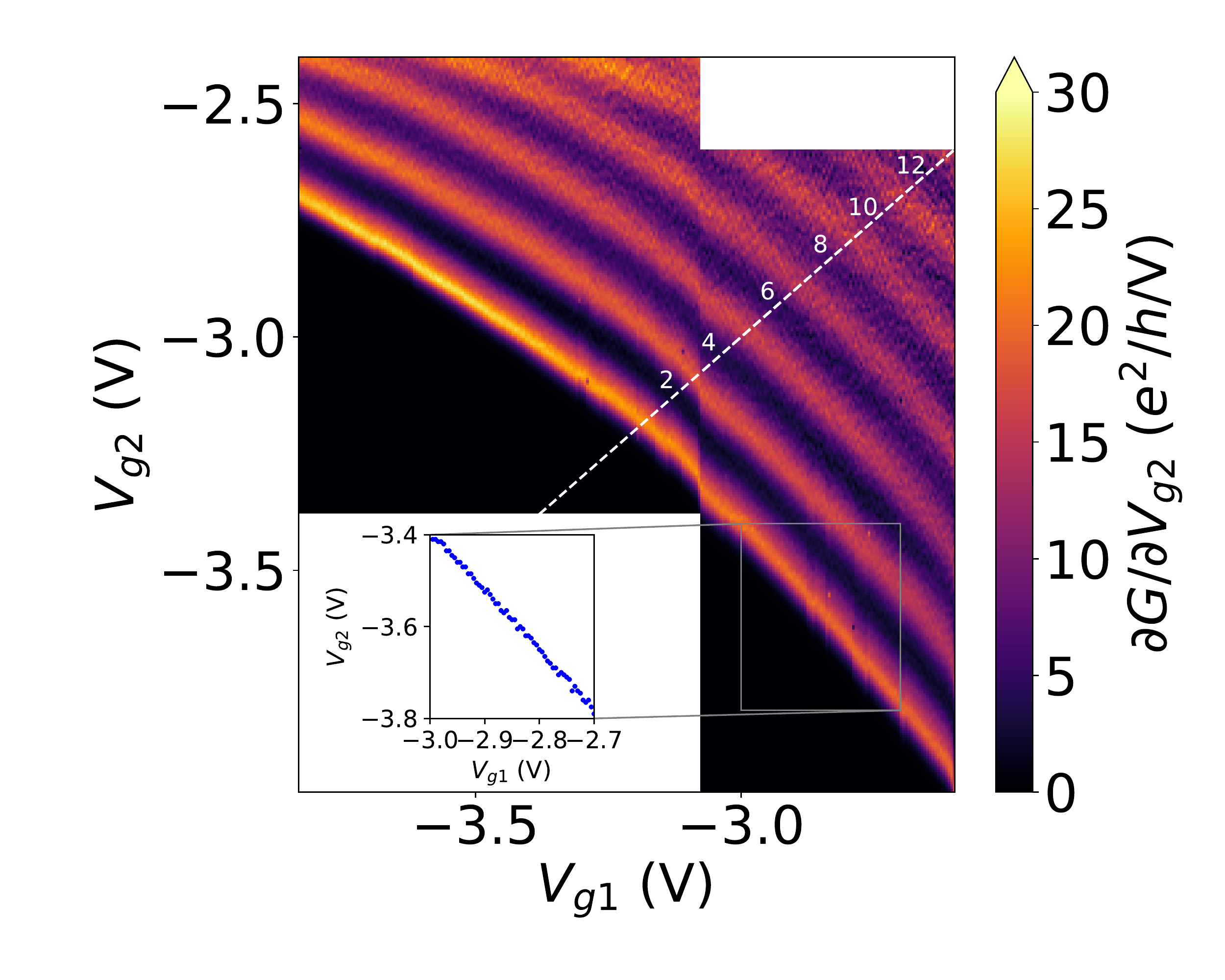}
\caption{Transconductance with respect to fast sweep axis $V_{\text{g2}}$, as a function of the two gate voltages, with data taken in two sweeps separated by a few hours. The discontinuity at $V_{\text{g1}} = -3.08$ V corresponds to drift in the QPC conductance between sweeps (see the SM \cite{SM} for discussion on stability).
The dark regions correspond to conductance plateaus, labeled in units of $e^2/h$, while the bright regions indicate transitions between them. In the shown gate voltage range, at least four conductance plateaus are visible. The white dotted line marks the trajectory of the symmetric gate sweep ($V_{\text{g1}}$ and $V_{\text{g2}}$) used in this work. The inset shows a fit to the first transconductance maxima in the $V_{\text{g1}} \in [-3,-2.7]$ V and $V_{\text{g2}} \in [-3.8,-3.4]$ V range, with discontinuities (at $V_{\text{g1}} = -2.84$ V for example) arising from a drift in the QPC conductance between traces along the slow sweep axis $V_{\text{g1}}$.} 
 	 \label{fig:asym_gate}
            \end{figure}

By applying an asymmetric voltage bias to the QPC split gates, we can laterally shift the position of the confining potential in real space. This serves as a spatial map of localized disorder or other potential fluctuations which may increase backscattering in the channel or create accidental quantum dots \cite{WilliamsonQPClateral}. Tuning the two gate voltages independently, the transconductance with respect to the fast sweep axis $V_{\text{g2}}$ is shown in Fig.~\ref{fig:asym_gate}. The bright features correspond to transitions between conductance plateaus, and they appear consistently smooth across the entire range. Resonances caused by localized disorder would appear as additional gate voltage-dependent lines in this map; the absence of such features here suggests a clean, defect-free channel within this range. Tuning the gate asymmetry to avoid spurious resonances is a common technique in QPC operation---not needing it here will significantly simplify the operation of devices with larger numbers of gates, where cross-capacitances must be diligently accounted for. The blue dots in the inset show a fit to the first transconductance peak in the $V_{\text{g1}} \in [-3,-2.7]$ V and $V_{\text{g2}} \in [-3.8,-3.4]$ V range. The discontinuities in the fit are due to a drift in the QPC conductance between traces along the slow sweep axis $V_{\text{g1}}$. We additionally note that we do not observe signatures of the 0.7 anomaly or of half-quantized plateaus at zero magnetic field in this QPC, though they have been reported previously in similar structures \cite{ShabaniQPC2014, LeeQPC2019, Matsuo2017}. This raises the question of the universality of such features in heterostructures of this type.

\section{Conclusion}
We have presented the fabrication and characterization of a QPC in an InAs-based deep quantum well which displays remarkable cleanliness despite the lattice-mismatched InP substrate. Transport through the QPC is smoothly quantized at zero and finite B-field, and bias spectroscopy reveals a harmonic confining potential with large subband spacing of near $5$ meV. We find an isotropic in-plane $g$ factor $|g_{x,y}^*| = 12 \pm 2$ and an out-of-plane $g$ factor $|g_{\perp}^*| = 27 \pm 1$.

This study supports the integration of QPCs as tunable tunnel barriers, charge sensors \cite{Elzerman2004, Reilly2007}, or mode collimators \cite{vanHouten1988TMF, Ji2003} into more complex InAs-based two-dimensional quantum devices such as quantum dots.
This is a critical building block toward investigations in spintronics, spin qubits, and hybrid superconductor-semiconductor topological physics.

The data that support the findings of this study are available from the corresponding author upon request.

\begin{acknowledgments}
We thank W. Pouse, E. Mikheev, J. Williams, M. Pendharkar, A.C.C. Drachmann, K. Ensslin, T. Ihn, Z. Lei, H. Sahasrabudhe, R. Rahman and F. Pierre for their scientific insights and suggestions. Measurement and analysis were supported by the U.S. Department of Energy (DOE), Office of Science, Basic Energy Sciences (BES), under Contract No. DE-AC02-76SF00515. Growth and characterization of heterostructures was supported by Microsoft Quantum. C.L.H. acknowledges support from the National Science Foundation (NSF) and Stanford Graduate Fellowship (SGF). Part of this work was performed at the Stanford Nano Shared Facilities (SNSF), supported by the National Science Foundation under Award No. ECCS-2026822.
\end{acknowledgments}

\pagebreak
\widetext
\begin{center}
\textbf{\large Supplemental Material: Clean quantum point contacts in an InAs quantum well grown on a lattice mismatched InP substrate}
\end{center}
\date{\today}
\setcounter{equation}{0}
\setcounter{figure}{0}
\setcounter{table}{0}
\setcounter{page}{1}
\makeatletter
\renewcommand{\theequation}{S\arabic{equation}}
\renewcommand{\thefigure}{S\arabic{figure}}
\renewcommand{\thesection}{S\arabic{section}}
\renewcommand{\bibnumfmt}[1]{[S#1]}
\renewcommand{ \citenumfont}[1]{S#1}

\setcounter{figure}{0}
\setcounter{section}{0}

\section{Indium Arsenide quantum well heterostructure stack}

\begin{figure}[!htbp] 
            \centering
            \includegraphics[width=\textwidth,keepaspectratio]{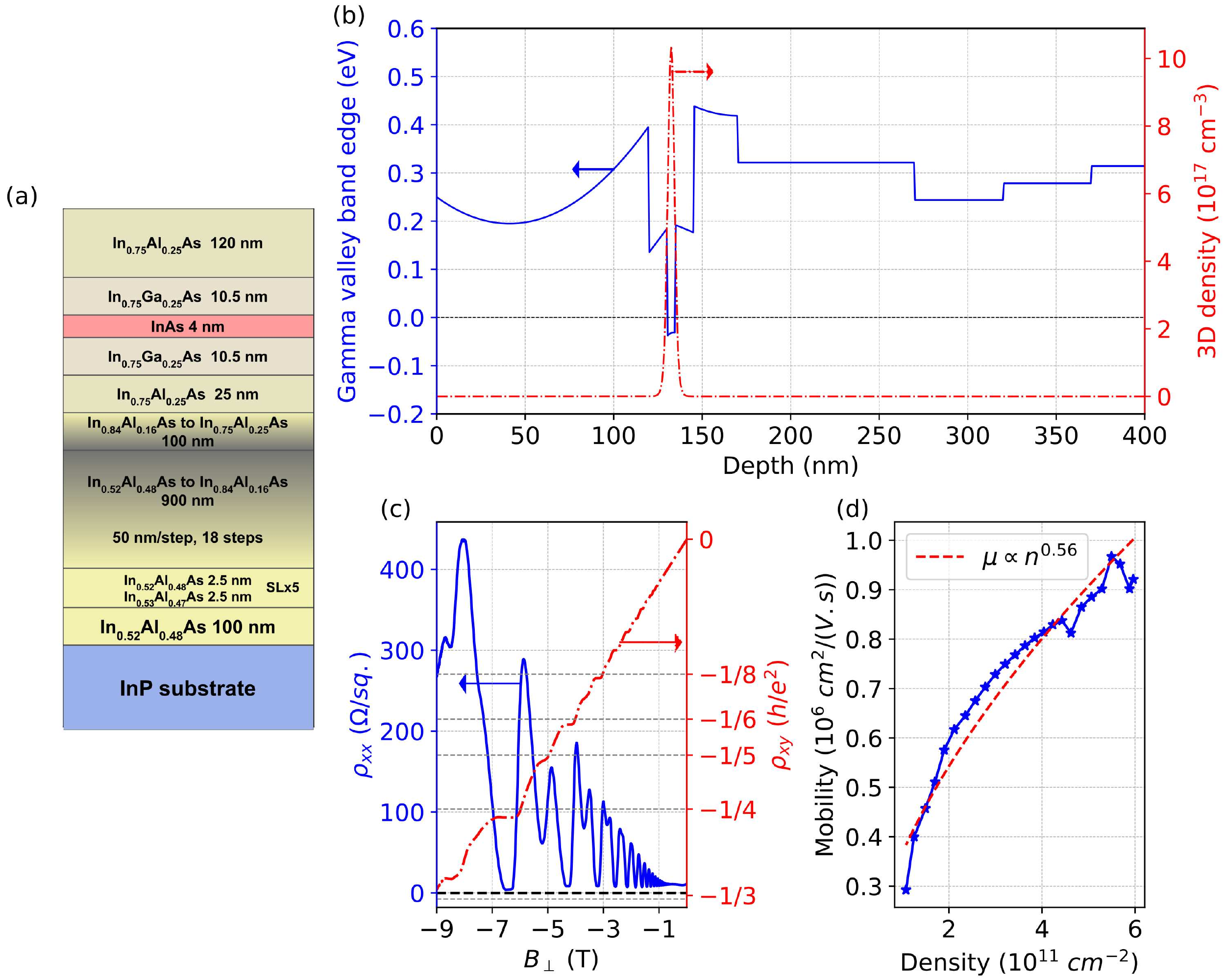}
\caption{(a) Cross-sectional schematic showing the layer structure of the InAs heterostructure stack. (b) $\Gamma$-valley conduction band edge and electron density of the quantum well from a self-consistent 1D Schr\"{o}dinger-Poisson simulation using NEMO5. The energy scale in (b) is referenced to the Fermi level. (c) Four-terminal magnetotransport data for a $40$ $\upmu$m wide Hall-bar, with a Ti/Au gate to tune the density. (d) Mobility variation with density, with a $\mu \propto n^{0.56}$ power law fit to the data. See text for discussion.} 
 	 \label{fig:stack}
            \end{figure}

A cross-sectional schematic of the InAs quantum well (QW) layer structure is shown in Fig.~\ref{fig:stack}(a). The 4 nm InAs QW sandwiched between 10.5 nm In$_{0.75}$Ga$_{0.25}$As barriers forms  the active region of the stack and hosts the two-dimensional electron gas (2DEG). Self-consistent Schr\"{o}dinger-Poisson simulations with NEMO5 \cite{NEMO5S} (see Fig.~\ref{fig:stack}(b)) reveal an electron density concentrated in the InAs QW, with a single subband occupied. The 2DEG electron density extracted from low-field Hall resistance and Shubnikov-de Haas (SdH) oscillations in a 40 $\upmu$m wide Hall-bar with a Ti/Au top-gate fabricated on the same chip corroborate the occupation of a single subband, with density $n_{\text{Hall}} \simeq n_{\text{SdH}} =  6\times 10^{11}$ cm$^{-2}$ and a mobility of $\mu = 9.3\times 10^{5}$ cm$^2$/Vs (see Fig.~\ref{fig:stack}(c)). 

The mobility variation with density probed by energizing the top-gate is depicted in Fig.~\ref{fig:stack}(d), with a fit to a $\mu \propto n^{\alpha}$ power-law. A best-fit exponent $\alpha \rightarrow 0.56$ indicates that the mobility is limited by unintentional background impurities and native charged point defects \cite{SDS_scaling2013S}.

\section{Additional fabrication details}
Electron-beam lithography was performed at the Stanford Nano Shared Facilities with a 100 kV JEOL-6300FS using 495 PMMA A5 resist and developed in 1:3 methyl isobutyl ketone (MIBK): isopropanol. Resist removal was aided by a remote oxygen plasma descumming step. The wet mesa etchant comprised 12.01 g of citric acid anhydrous in solution with 250 mL H\textsubscript{2}O, 3 mL H\textsubscript{3}PO\textsubscript{4} (85\%), and 3 mL H\textsubscript{2}O\textsubscript{2} (30\%). The etch depth of 300 nm was chosen to avoid parallel conduction in the buffer layer which had been observed in similar heterostructures but not necessarily this one. In the ohmic contact step, following the mask development the sample was dipped in a batch of the mesa etchant (as described above) for 15 seconds to remove surface layers and additionally was \textit{in situ} milled for 15 seconds in a gentle Ar plasma directly prior to metallization. The HfO\textsubscript{2} dielectric layer was grown by thermal ALD using a Cambridge Nanotech (now Veeco) Savannah.             

\section{Instrumentation and Measurement Details}
Voltage-biased transport measurement were performed as shown in the schematic of the measurement setup in Fig.~1. A low frequency ($<20$ Hz) ac  1 V rms sinusoidal signal was sourced from a Stanford Research Systems SR830 lock-in amplifier. The variable dc bias voltage $V_{\text{bias}}$ was provided by a Keithley 2400. The ac  and dc signals were scaled and summed using a resistive voltage adder network, with scaling factors 1E-4 and 1E-2 respectively. The summed voltage bias signal was applied to the source terminal of the extended Hall-bar like mesa, while the drain terminal was connected to a virtual ground provided by an Ithaco 1211 current preamplifier with a gain setting $10^7$ V/A ($10^6$ V/A) for zero (finite) dc bias. The ac  output of the current amplifier, $I_{\text{ac}}$ was measured using a Stanford Research Systems SR830 lock-in amplifier, and the dc output $I_{\text{dc}}$ was measured using an Agilent 34401A digital multimeter. The voltage drop, measured diagonally across the QPC, was differentially amplified by a factor of 100 using a Stanford Research Systems SR560 voltage preamplifier.  The dc output of the preamplifier, $V_{\text{dc}}$, was measured using an Agilent 34401A digital multimeter, and the ac  output, $V_{\text{ac}}$, was measured using a Stanford Research Systems SR830 lock-in amplifier. The QPC gate voltages were sourced from a pair of Keithley 2400s. 

\section{Effective mass estimation from finite-bias spectroscopy}
The electron effective mass $m^*$ determines the cyclotron frequency $\omega_c = e B_{\perp}/m^*$ in an out-of-plane magnetic field $B_{\perp}$, and can be probed by applying a finite dc bias across the source-drain terminals of the Hall-bar, as described in Sec.~III.\ 2 of the main paper. The extent along $V_{dc}$ of the transconductance diamond $(\Delta E_2)$ corresponding to the quantized conductance $G=2e^2/h$ can be expressed as 
\begin{equation}
    \Delta E_{2} = \hbar\sqrt{\omega_y^2 + \omega_c^2} - \frac{1}{2}g^*_{\perp,1}\mu_BB_{\perp}- \frac{1}{2}g^*_{\perp,2}\mu_BB_{\perp}, 
    \label{eq:effmass}
\end{equation}
where $\hbar \omega_y$ is the gate voltage-dependent QPC confinement energy, and $g^*_{\perp,n}$ is the effective out-of-plane $g$ factor for the $n^{\text{th}}$ spin-degenerate subband.

\begin{figure}[!htbp] 
            \centering
            \includegraphics[width=0.8\textwidth,keepaspectratio]{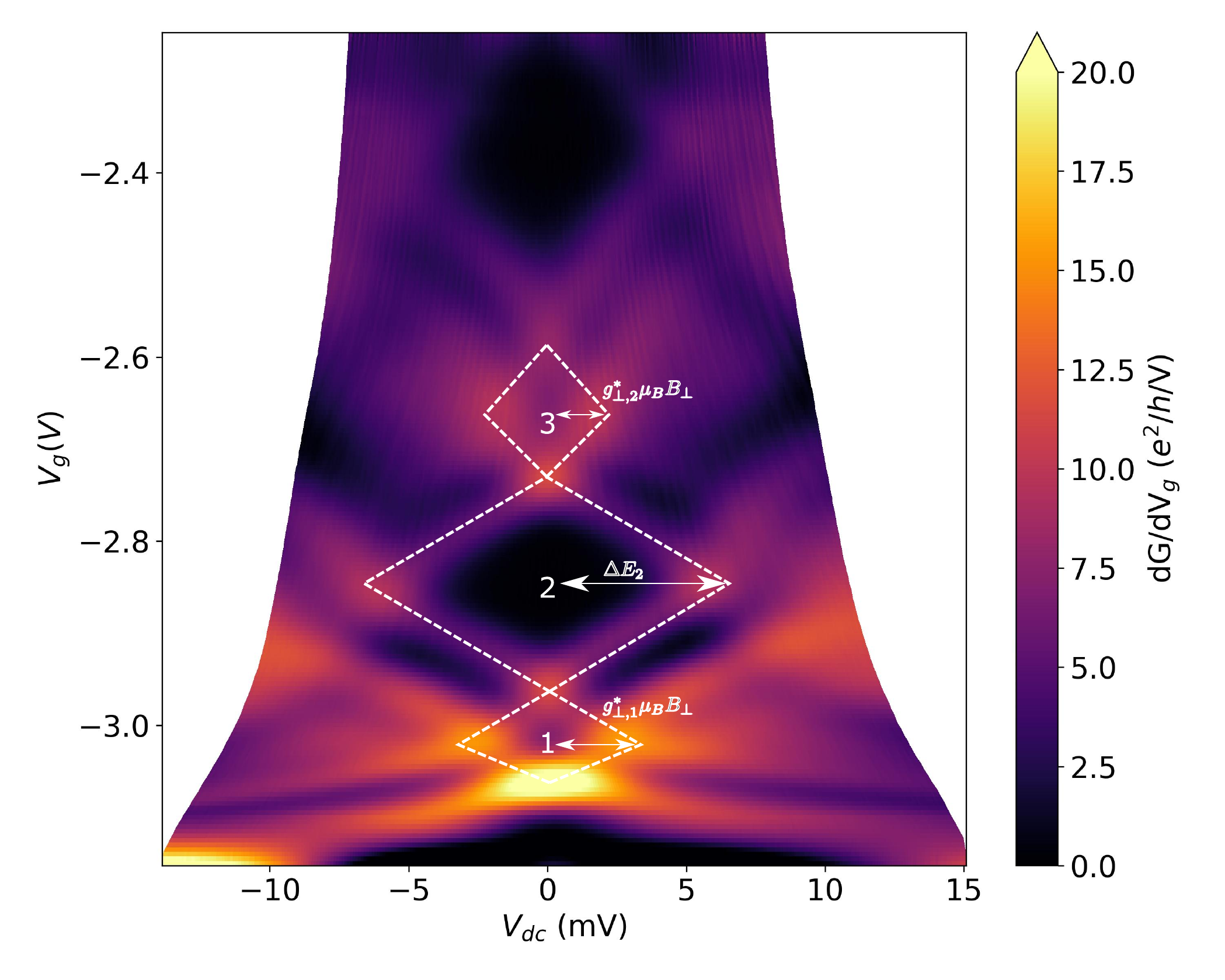}
\caption{Transconductance $dG/dV_{\text{g}}$ as a function of \Vg and $V_{\text{dc}}$ at $B_{\perp}$ = 1.983 T. The white dashed lines highlight the quantized conductance plateaus, labeled in units of $e^2/h$. The extent of the diamonds along $V_{\text{dc}}$ measure the spin-split subband spacings, governed by Eq.~\ref{eq:effmass}.} 
 	 \label{fig:effmass}
            \end{figure}

Figure~\ref{fig:effmass} shows the transconductance as a function of \Vg and $V_{\text{dc}}$ at $B_{\perp} = 1.983$ T. This results in a bulk filling factor $\nu_{\text{bulk}} = 10$. The diamonds corresponding to $G=1e^2/h, 2e^2/h$ and $3e^2/h$ are highlighted with white dashed lines. From the $V_{\text{dc}}$ extent of the diamonds we extract $g^*_{\perp,1}\mu_BB_{\perp} = 3$ meV, $g^*_{\perp,2}\mu_BB_{\perp} = 2$ meV, and $\Delta E_2 = 6$ meV. Taking $\hbar \omega_y \sim 5$ meV at \Vg $\sim -2.83$ V (recall Fig.~2(d)), we extract $\hbar \omega_c = 6.9$ meV at $B_{\perp} = 1.983$ T corresponding to $m^* = 0.033m_e$.

\section{Saddle-point potential model}
Given the near-harmonic confinement potential (Sec.~III.\ 2), we apply B\"{u}ttiker's saddle-point potential model to extract the confinement potential parameters from our linear conductance data. The saddle-point model extends the transverse harmonic confinement potential to the 2D transport plane by adding a parabolic drop-off in the longitudinal direction. This results in a broadening of the conductance transitions, while maintaining the harmonic subband spectra $\hbar \omega_y\left(n-\frac{1}{2}\right)$. The potential has a saddle-point at the center of the constriction $(x=y=0)$, and can be expressed as
\begin{equation}
    V(x,y) = V_0 - \frac{1}{2}m^*\omega_x^2x^2 + \frac{1}{2}m^*\omega_y^2y^2,
\end{equation}
where $V_0$ is the potential at the center of the constriction, and $\omega_{x(y)}$ parameterizes the longitudinal (transverse) harmonic potential. The transmission probability for mode $n$ at the Fermi energy $E_F$ is given by 
\begin{equation}
    T_n(E_F) = \frac{1}{1 + \exp \left(-2\pi \epsilon_n/\hbar \omega_x\right)},
\end{equation}
where $\epsilon_n = E_F - E_0 - \hbar\omega_y\left(n-\frac{1}{2}\right)$, and $E_0$ is the conduction band edge. The linear response conductance 
can then be expressed as
\begin{equation}
    G = g_s\frac{e^2}{h}\int_{-\infty}^{\infty}dE\sum_nT_n(E)\left(-\frac{\partial f}{\partial E}\right),
    \label{eq:spG}
\end{equation}
where $g_s = \{1,2\}$ represents the subband degeneracy in the presence and absence of a magnetic field respectively, and $f(E) = 1/\left(1 + \exp\left(\frac{E}{k_BT}\right)\right)$ is the Fermi function. Conductance plateaus are well resolved in the limit $\omega_y/\omega_x \gg 1$.

\begin{figure}[!htbp] 
            \centering
            \includegraphics[width=\textwidth,keepaspectratio]{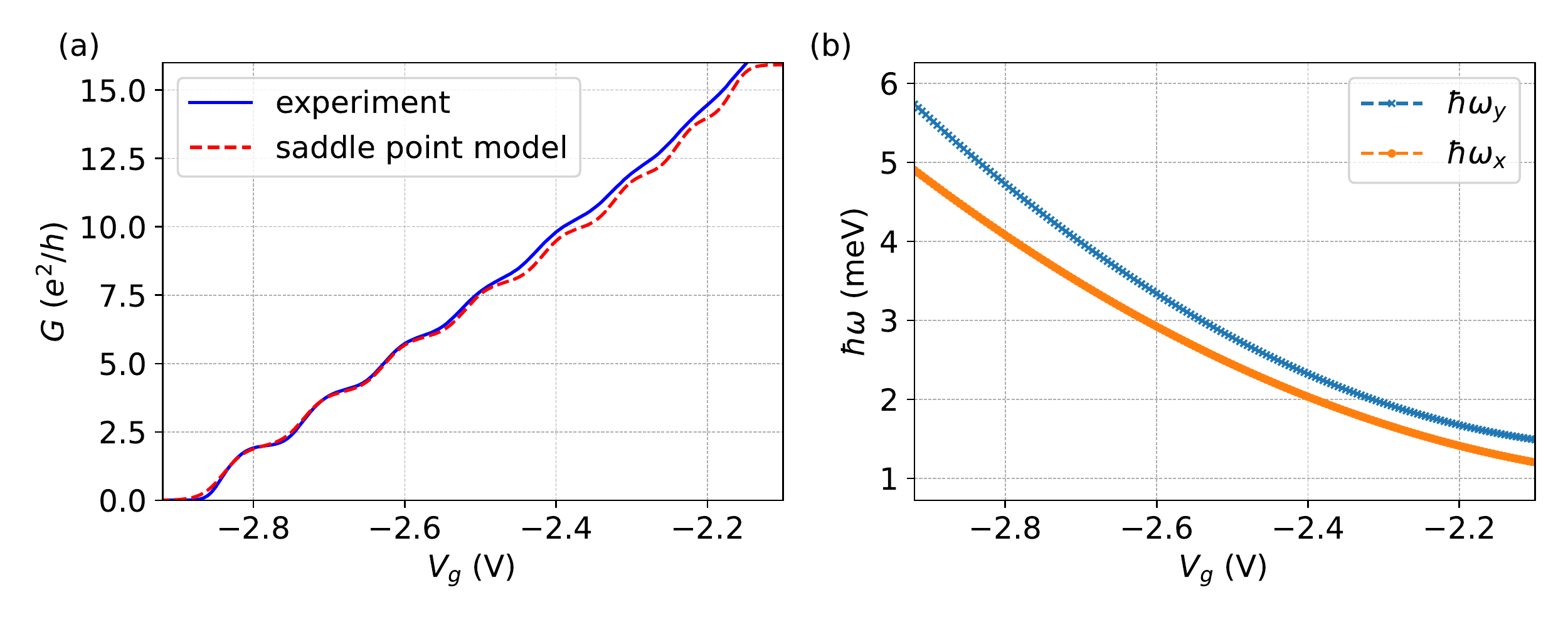}
\caption{(a) The conductance computed from a saddle-point potential model (red, dashed) fit to the measured conductance of the QPC (blue, continuous).  A good fit to the experiment is observed in the few-mode limit $(G \lesssim 8e^2/h)$. See text for discussion. (b) The transverse confinement parameter $\hbar\omega_y$ is estimated from finite-bias spectroscopy at zero magnetic field, with the longitudinal broadening parameter $\hbar\omega_x$ serving as the fit parameter. A quadratic dependence of $\hbar\omega_{x,y}$ on \Vg is assumed.} 
 	 \label{fig:sp}
            \end{figure}

Finite-bias spectroscopy at zero field gives an estimate of the transverse confinement potential parameter $\hbar \omega_y$ from the variation of the subband spacings with \Vg (recall Fig.~2(d) in the main text). A good fit to a quadratic dependence of $\hbar\omega_y$ on \Vg is observed. The broadening parameter $\hbar\omega_x$ can be estimated by fitting the measured conductance at zero magnetic field to Eq.~\ref{eq:spG}. Assuming a quadratic dependence of $\hbar \omega_x$ on \Vg, Fig.~\ref{fig:sp} shows the best fit (red) to the experimental curve (blue). We infer that a saddle-point potential is a good description of the constriction in the few-mode limit $(G \leq 8e^2/h)$. The error in estimating the subband spacing in the many-mode limit may limit the applicability of the model for \Vg $>-2.4 $V. Furthermore, as seen in GaAs QPCs \cite{Geier2020S} the higher constriction density as the QPC is opened up leads to increased screening of the confinement potential, resulting in the transition from a harmonic confinement towards a flat-bottom potential well \cite{LAUX1988101S}. 
\section{Gate voltage dependence of constriction width}\label{sec:WvsVg}
\begin{figure}[!htbp] 
            \centering
            \includegraphics[width=0.8\textwidth,keepaspectratio]{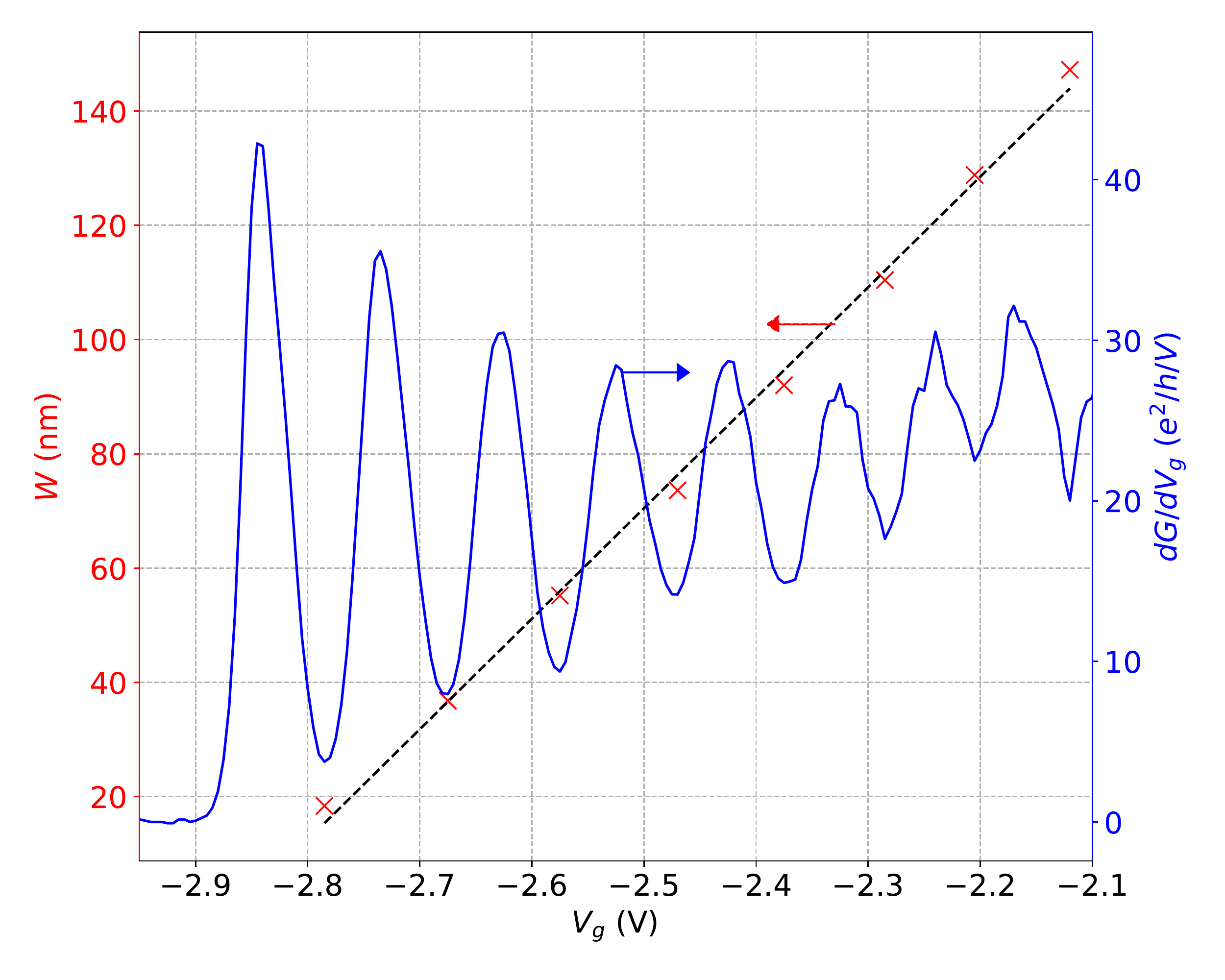}
\caption{The constriction width extracted from conductance data at zero field. The slope of the black dashed line is $193 \pm 4$ nm/V.} 
 	 \label{fig:WvsVg}
            \end{figure}

The constriction width $W_{\text{qpc}}(V_g)$ as a function \Vg can be extracted from the conductance data at zero field, by recognizing $W_{\text{qpc}}(V_g) = n\times \lambda_F/2$ at the $n$-th transconductance minima, where $\lambda_F/2$ is the Fermi wavelength in the 2DEG. Figure~\ref{fig:WvsVg} shows the estimated $W_{\text{qpc}}(V_{\text{g}})$ which fits a straight line with slope $193 \pm 4$ nm/V. 

\section{QPC split-gate lever arm}\label{sec:alpha}
The capacitive lever arm of the QPC gives the change in Fermi energy for 1 V change in \Vg. In other words, the lever arm quantifies the capacitive coupling between the split-gates and the constriction. Section~III.\ 2 in the main text describes finite-bias spectroscopy on the QPC. The lever arm $\alpha$ can be extracted as half the absolute slope of the transconductance maxima 
\begin{equation}
    \alpha = \frac{1}{2}\left|\frac{dV_{\text{dc}}}{dV_{\text{g}}}\right|,
    \label{eq:alpha}
\end{equation}
since half the diamond width along the $V_{\text{dc}}$ axis corresponds to an energy scale equivalent to the full diamond height along the \Vg axis. 

The chemical potential $\mu(V_{\text{g}})$, defined as the Fermi level offset w.r.t. the conduction band edge can be expressed as
\begin{equation}
    \mu(V_{\text{g}}) = \int_{V_{\text{po}}}^{V_{\text{g}}} \alpha\left(V'_\text{g}\right) d{V'_\text{g}},
    \label{eq:mu}
\end{equation}
where $\mu = 0$ at the pinch-off gate voltage $V_{\text{g}} = V_{\text{po}}$. 

To translate the subband spin-splittings in an in-plane magnetic field from gate voltage (Fig.~3(a)) to chemical potential (Fig.~3(b)), we perform the following steps --
\begin{enumerate}
    \item Estimate the lever-arm $\alpha_0$ from finite-bias measurements of Fig.~2(d) as described by Eq.~\ref{eq:alpha}.
    \item Note that the zero-field pinch-off gate voltage shifted from -2.845 V to -3.184 V between the finite-bias measurements in Fig.~2(d) and in-plane field measurements in Fig.~3(a). Offset the lever-arm $\alpha_0$ to include for this conductance drift, $\alpha = \alpha_0({V_\text{g}} + 0.339)$, and fit a cubic polynomial to $\alpha$.
    \item Using Eq.~\ref{eq:mu} and $V_{\text{po}} = -3.184$ V, evaluate $\mu(V_{\text{g}})$.
\end{enumerate}

\begin{figure}[!htbp] 
            \centering
            \includegraphics[width=0.8\textwidth,keepaspectratio]{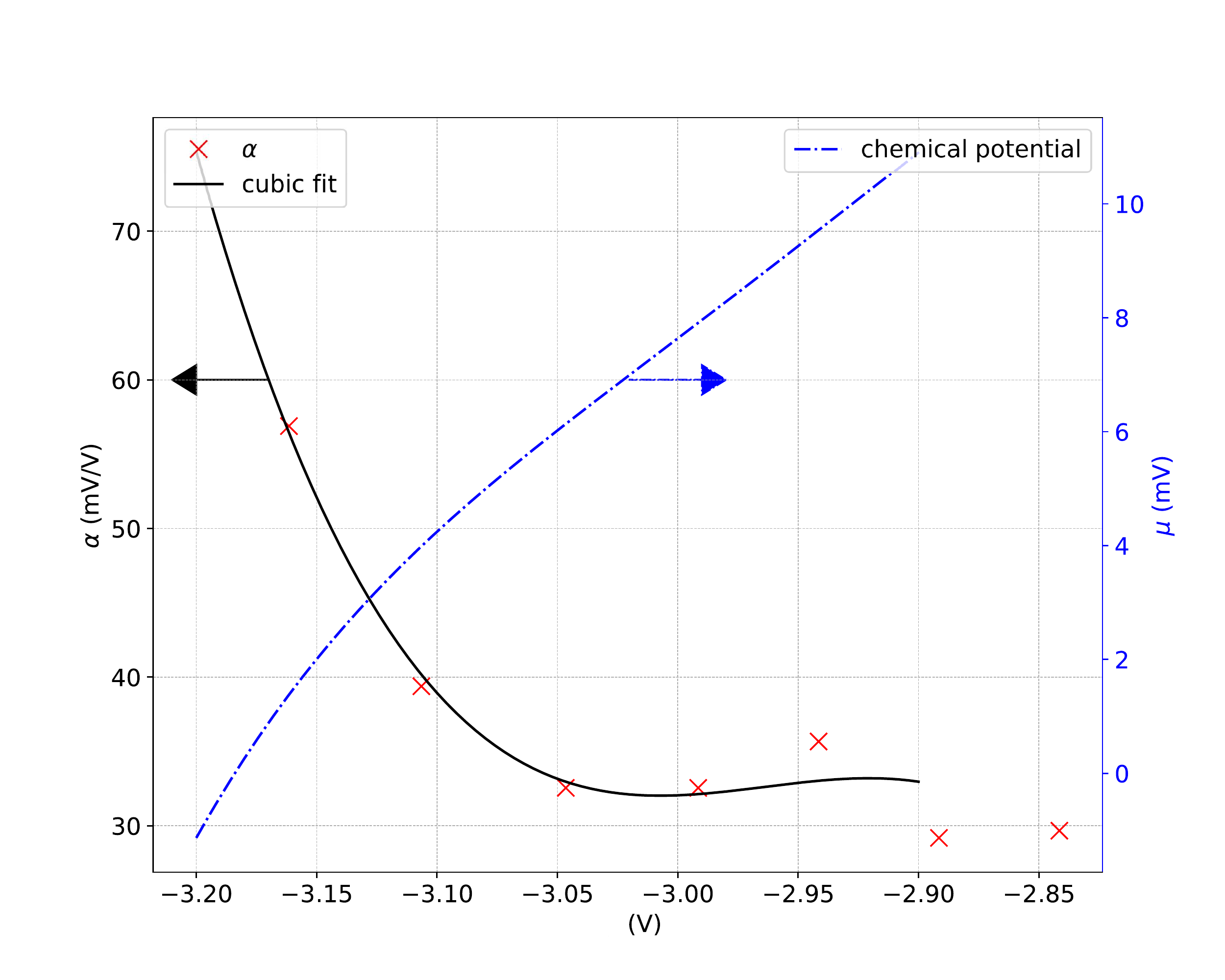}
\caption{The split-gate lever arm $\alpha$ (red cross) as a function of \Vg, extracted from the slope of the transconductance peaks in a finite dc bias measurement at zero field and offset by -0.339 V to include the conductance drift between Fig.~2(d) and Fig.~3(a). An expected decrease is observed as the constriction gets wider. The black solid curve shows a cubic polynomial fit to $\alpha$. The blue broken curve shows the chemical potential $\mu(V_{\text{g}})$ evaluated using Eq.~\ref{eq:mu}.} 
 	 \label{fig:alpha}
            \end{figure}

Figure~\ref{fig:alpha} plots the lever arm $\alpha$,  from the first three spin-degenerate conductance plateau diamonds of Fig.~2(d), with a -0.339 V offset to include the conductance drift. Also shown is the cubic polynomial fit to $\alpha$, and the evaluated $\mu(V_{\text{g}})$. The broadening of the transconductance peaks makes a precise estimation of $\alpha$ challenging. We observe an expected decrease in $\alpha$ as the constriction gets wider and the capacitive coupling reduces. 

\section{In-plane magnetic field perpendicular to transport direction}\label{sec:Bpar_y}
\begin{figure}[!htbp] 
            \centering
            \includegraphics[width=\textwidth,keepaspectratio]{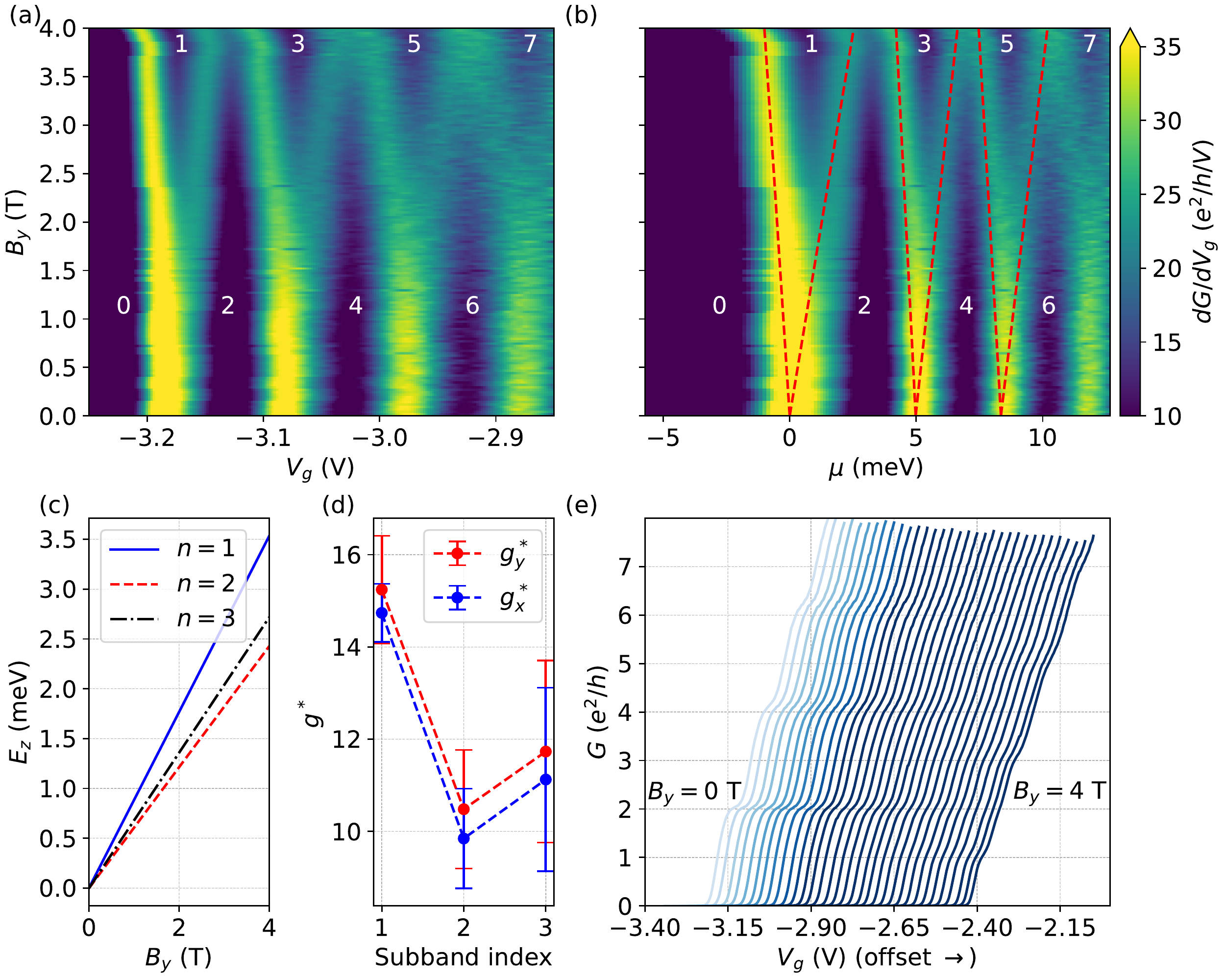}
\caption{In-plane magnetic field spectroscopy showing (a) transconductance $dG/dV_{\text{g}}$ as a function of QPC gate voltage $V_{\text{g}}$ and a magnetic field $B_{y}$ applied in the plane of the sample and perpendicular to the transport direction, and (b) as a function of chemical potential $\mu$ estimated from the capacitive lever arm (see Sec.~\ref{sec:alpha}). The dark regions correspond to conductance plateaus labeled in units of $e^2/h$. The transitions between conductance plateaus are visible as bright regions. The appearance of additional dark regions at high $B_{y} (\gtrsim 3 \text{ T})$ is a signature of a Zeeman-induced spin-splitting of the subbands. (c) The Zeeman energy for the first three subbands, extracted from a linear fit to the spin-split transitions (red dotted lines in (b)). (d) The effective in-plane $g$ factor parallel ($g^*_x$) and perpendicular ($g^*_y$) to the transport direction estimated from the slopes of the Zeeman energy in (c) for the first three subbands. Within the error bars, the in-plane effective $g$ factor is isotropic and close to the bulk value for InAs $|g|=13$.
(e) Conductance as a function of $V_{\text{g}}$ at various fixed $B_{y}$. The traces in (e) are offset along the horizontal axis for clarity and display a progressive development of conductance plateaus at $1e^2/h$, $3e^2/h$ and $5e^2/h$.}
 	 \label{fig:Bpar_y}
\end{figure}

The effective $g$ factor in-plane but perpendicular to the transport direction $g^*_y$ can be extracted by repeating the measurement described in Sec.~III.\ 3 with a magnetic field $B_y$ in the $y$-direction (see Fig.~1 for axes orientation). The transconductance as a function of $B_y$ and \Vg is shown in Fig.~\ref{fig:Bpar_y}(a), with dark regions corresponding to conductance plateaus labeled in units of $e^2/h$. Similar to the $B_x$ data from Fig.~3(a) in the main text, the Zeeman effect of the applied field breaks spin degeneracy and gives rise to odd-integer plateaus. The transconductance has local maxima when the source/drain chemical potential is in resonance with a subband edge. Conductance traces for $B_y \in [0,4]$ T is shown in Fig.~\ref{fig:Bpar_y}(e), with the expected emergence of conductance plateaus at $1e^2/h$ and $3e^2/h$ as $B_y$ is increased, and the corresponding decrease in the width of the even-integer plateaus.

As described in the main text and Sec.~\ref{sec:alpha}, the spin-splitting of the subbands is further elucidated in Fig.~\ref{fig:Bpar_y}(b) by translating \Vg to chemical potential $\mu$, using the split-gate lever-arm $\alpha = d\mu/dV_{\text{g}}$. A linear fit to the transconductance maxima, constrained to intersect at $B_y=0$ for each spin-split subband pair, is used to the extract the Zeeman energy $E_Z$ as a function of $B_y$, as depicted in Fig.~\ref{fig:Bpar_y}(c). The in-plane $g$ factor extracted from the slope of the Zeeman energy is shown in Fig.~\ref{fig:Bpar_y}(d), revealing negligible anisotropy between $x$ and $y$ directions.  The  error  estimates  are based on  fitting  parameter  variances. See Sec.~III.\ 4 in the main text for further discussion.

\section{Out-of-plane $\mathbf{g}$-factor in the 2DEG}
The out-of-plane effective $g$ factor $g^*_{\perp}$ in the 2DEG can be extracted from the magnetoresistance measured in a Hall-bar with $B_{\perp}$. The many-body exchange enhancement to the $g$ factor has its origins in the spin-population difference of Landau-levels. The enhancement depends on the Fermi level position, and is maximized at odd filling factors -- which corresponds to the largest difference in spin-population. Furthermore, this many-body interaction is screened with increasing density, and consequently, a reduction in the $g_{\perp}^*$ oscillation amplitude is expected for higher filling factors. 

\begin{figure}[!htbp] 
            \centering
            \includegraphics[width=0.8\textwidth,keepaspectratio]{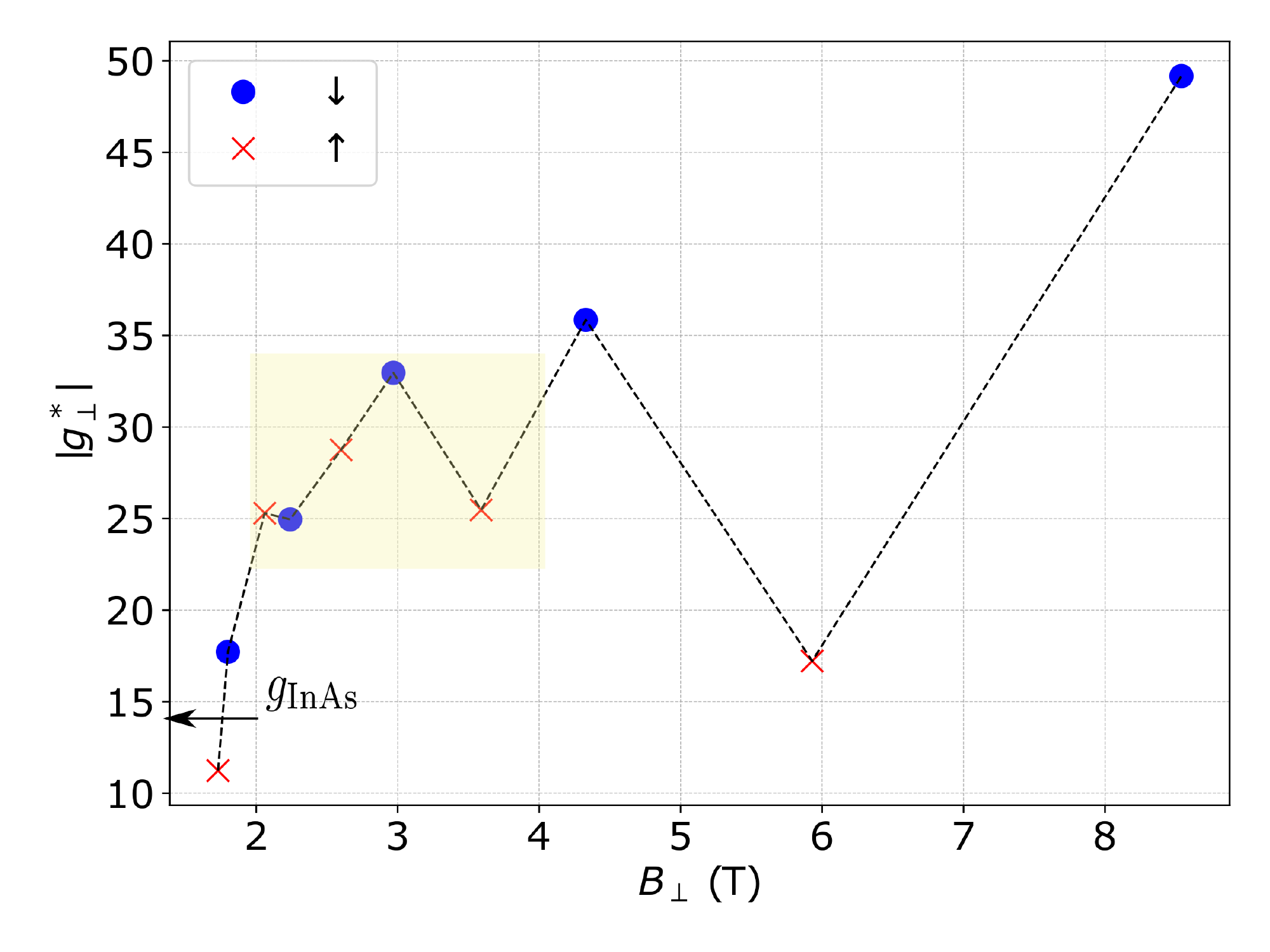}
\caption{The 2DEG effective $g$ factor extracted from the $R_{xx}$ maxima in a Hall-bar geometry. The successive depopulation of spin-split Landau levels (blue stars : up-spin, red crosses : down-spin) with $B_{\perp}$ leads to an oscillatory many-body exchange enhancement of $g^*_{\perp}$. The magnitude of the enhancement reduces with field, as the higher filling factors result in more efficient screening of the many-body exchange interaction. For $B_{\perp} \lesssim 2$ T (filling factor $\nu_{\text{2DEG}} > 10$), $g^*_{\perp}$ is close to the bulk InAs $g$ factor $\simeq 15$. The yellow highlighted box marks the field range used for the 1D $g^*_{\perp}$ estimation using finite-bias spectroscopy in the QPC (Sec.~III.\ 4).} 
 	 \label{fig:gfactor_2DEG}
            \end{figure}

Our approach for extracting $g^*_{\perp}$ is described in detail in Ref.~\onlinecite{RAYMOND1985271S}. The four-terminal longitudinal resistance $R_{xx}$ has a maximum whenever the Fermi level $E_F$ is in resonance with a spin-split Landau level. For a spin-split Landau level pair we have,
\begin{equation}
    E_F - E_0 = \frac{\hbar^2\pi n_s}{m^*}= \left(N_L-\frac{1}{2}\right)\hbar\omega_c \pm \frac{1}{2}g^*_{\perp}\mu_B \tilde{B}_{\perp},
    \label{eq:g2DEG}
\end{equation}
where $E_0$ is the bottom of the conduction band, $n_s$ is the 2DEG electron density, $N_L=\{1,2,\dots\}$ is the Landau level index, and $\tilde{B}_{\perp}$ is the out-of-plane magnetic field corresponding to each $R_{xx}$ maxima. The $g^*_{\perp}$ extracted from the measured magnetoresistance  (Fig.~\ref{fig:stack}(c)) and Eq.~\ref{eq:g2DEG} is depicted in Fig.~\ref{fig:gfactor_2DEG}. As discussed above, an oscillatory enhancement with successive depopulation of spin-split Landau levels is observed. The amplitude and mean value of the oscillations decreases towards the bulk value $|g|\simeq 15$ with increasing filling factor (decreasing field). The gigantic oscillations at low filling factor which decrease with field is strong evidence for many-body exchange interaction driven enhancement of $g^*_{\perp}$. Interestingly, in the $B_{\perp} \in [2,4]$ T  range (yellow box in Fig.~\ref{fig:gfactor_2DEG}) used to study the 1D $g^*_{\perp}$ in the QPC (recall Fig.~4(d)) the enhanced $g$ factor of the 2DEG $\simeq 30$. This implies that the enhanced $g^*_{\perp}$ in the constriction, and the large $g^*_{\perp,x}$ anisotropy (Sec.~III.\ 3, D) are driven by many-body interactions in the 2DEG.

\section{Notes on QPC stability and reproducibility}
In this section we present data from various QPCs fabricated on the same heterostructure stack. The measurements were made over 9 cooldowns, intermittently over a period of 11 months. We focus on hysteresis and drift with \Vg fixed as well as bidirectionally swept over a range as the figures of merit. 

\begin{figure}[!htbp] 
            \centering
            \includegraphics[width=1\textwidth,keepaspectratio]{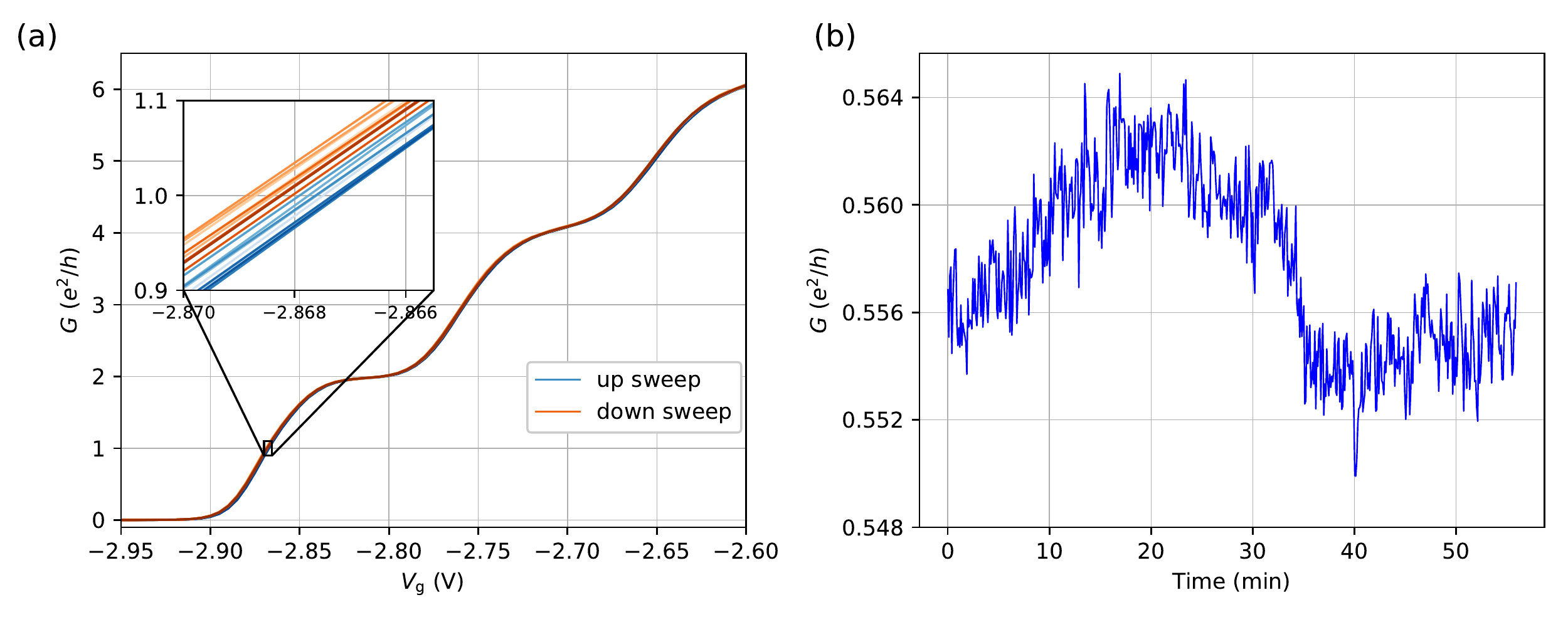}
\caption{(a) Conductance plotted as the QPC was swept back-and-forth in the range \Vg $\in [-3.13, -2.13]$ V ten times with a 5 mV step size. Inset: zoom of the curves at the middle of the first riser, highlighting that there was a hysteresis of $\sim 1$ mV between up and down direction, and curves swept in the same direction were reproducible within a 1 mV horizontal translation. Curves are ordered from light to dark trace. (b) The QPC was set along the first riser, and the conductance was watched for an hour, demonstrating conductance stability within 1.5\%.} 
 	 \label{fig:Stabilitygood}
            \end{figure}

Figure~\ref{fig:Stabilitygood}(a) tracks the conductance through the QPC (the same QPC as in the main text) as the symmetric gate voltage was swept bidirectionally ten times in the range \Vg $\in [-3.13, -2.13]$ V. Sweeps in the same direction are ordered from light to dark trace. Hysteresis between up and down sweeps was $\sim 1$ mV and sweeps in the same direction fell within a 1 mV horizontal translation of one another. In Fig.~\ref{fig:Stabilitygood}(b), the QPC was set to a position along the first riser ($G \sim 0.56$ $e^2/h$) and the device conductance was watched for one hour. The variation in conductance remained within 1.5\%. The stability in device operation allowed for longer sweeps such as the dc bias spectroscopy maps to be taken without additional tuning. 

\begin{figure}[!htbp] 
            \centering
            \includegraphics[width=0.45\textwidth,keepaspectratio]{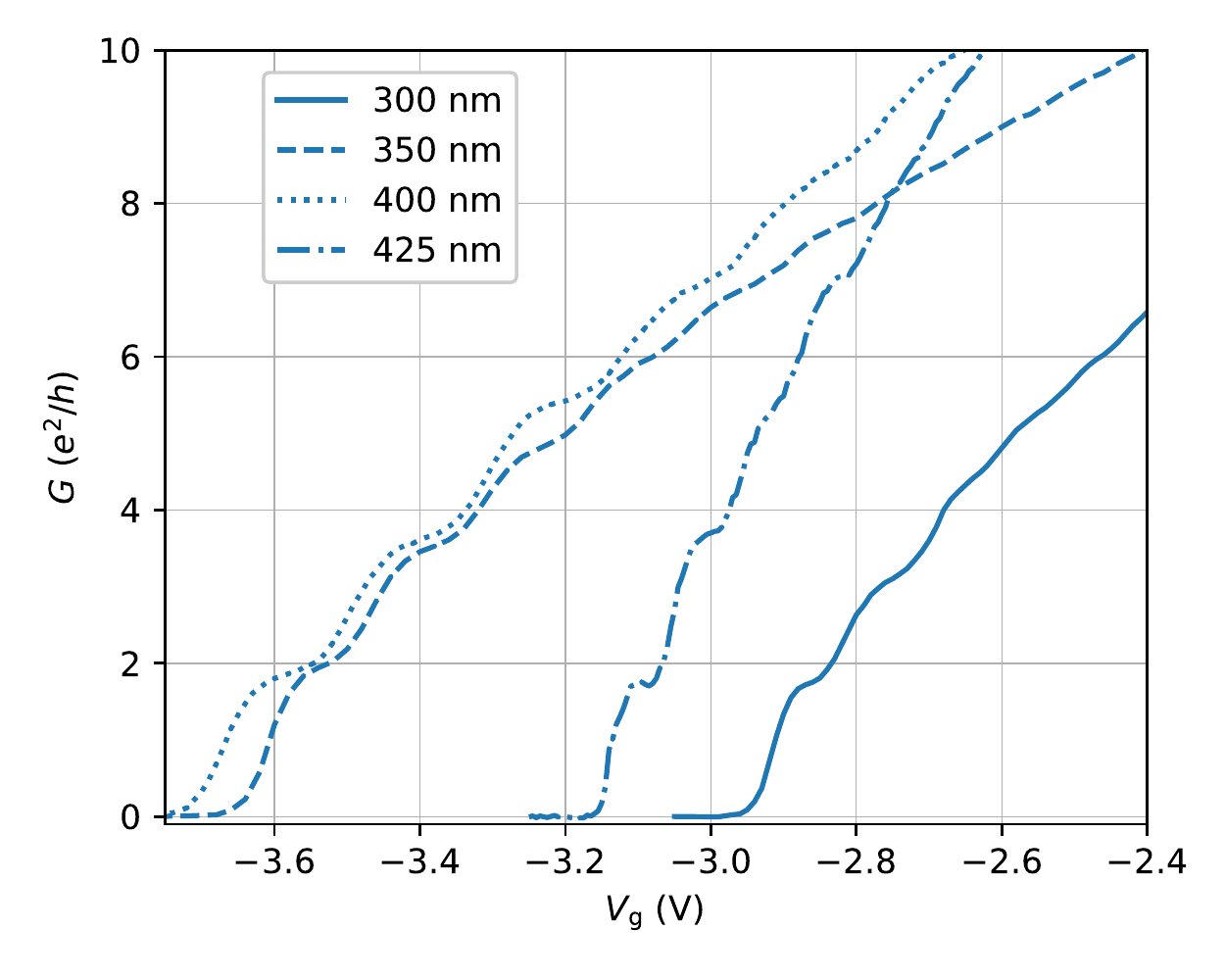}
\caption{Conductance data (taken at 1.5 K) from four additional QPCs, labeled by their designed gate separation. Curves are uncorrected for series resistance. The appearance of multiple steps in conductance is reproducible, though there is variability in the pinch-off voltage, the width and flatness of the plateaus, and the steepness of the risers.} 
 	 \label{fig:otherQPCs}
            \end{figure}

\begin{figure}[!htbp] 
            \centering
            \includegraphics[width=1\textwidth,keepaspectratio]{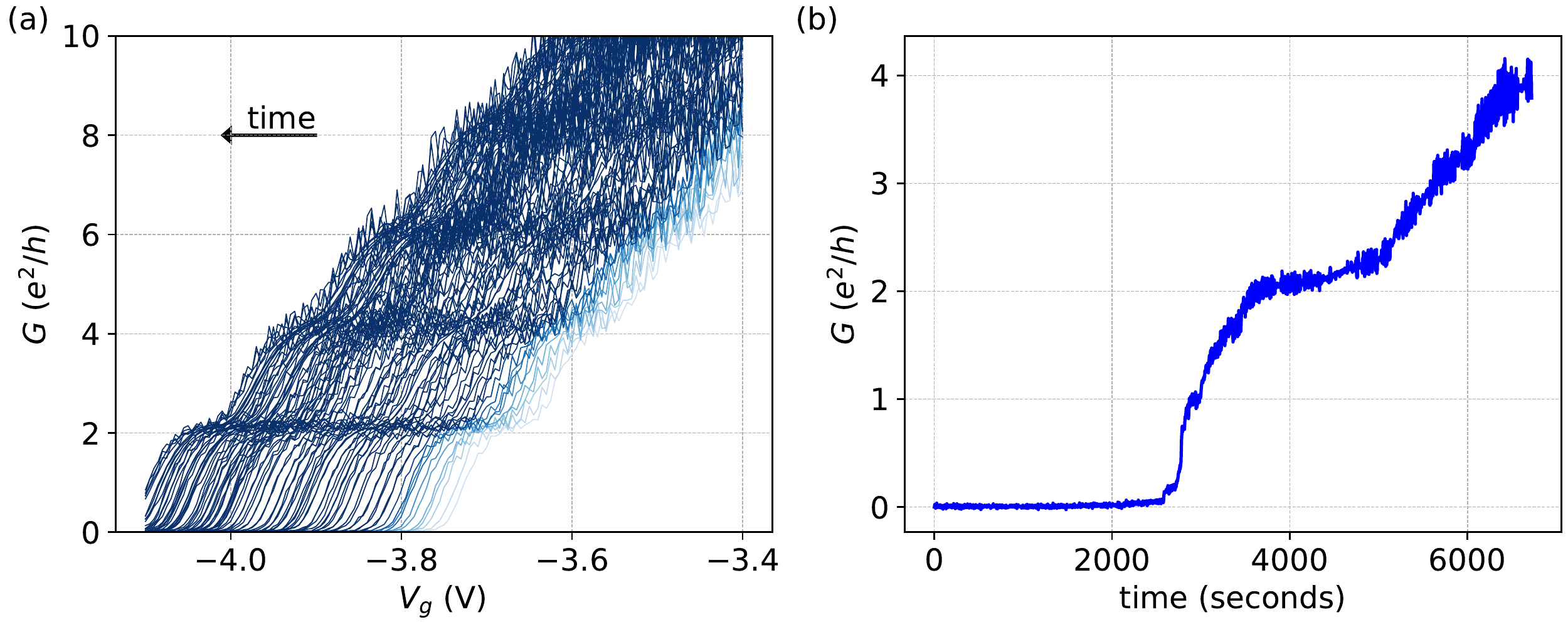}
\caption{Conductance data from the QPC featured in the paper, but taken over a previous cooldown. (a) The gates were swept in both directions $\in [-4.1,-3.4]$ V with a 5 mV step size. The pinch-off showed a continuous drift towards increasingly negative \Vg. Over the measurement duration of 4.5 hours, the pinch-off drifted by $> 400$ mV, showing no signs of stabilizing. (b) The conductance over a period of 2 hours with \Vg fixed at -3.38 V, prior to the procedure described in (a). Pinch-off lasted for 45 minutes before the conductance steadily grew to trace out a typical $G($\Vg$)$ curve.} 
 	 \label{fig:Stability2}
            \end{figure}

Conductance curves from four additional QPCs---fabricated on other mesas in the same fabrication run and measured over various cooldowns---are shown in Fig.~\ref{fig:otherQPCs}. Multiple conductance plateaus are consistently seen in these constrictions, although there are variations in pinch-off voltage, width and flatness of the plateaus, and steepness of the risers. A frequent failure mode for our QPCs was leakage between the gates and the 2DEG. In addition, some QPCs showed cooldown-to-cooldown variation, especially in conductance stability versus time. Figure~\ref{fig:Stability2}(a) shows conductance data from a separate cooldown from the same QPC as Fig.~\ref{fig:Stabilitygood}(a) and the main text while \Vg is swept bidirectionally in the range $[-4.1,-3.4]$ V. There is a persistent drift toward negative gate voltage with no indications of stabilizing. Figure~\ref{fig:Stability2}(b) shows the conductance through the QPC with \Vg fixed at $-3.38$ V, which begins in the pinch-off regime but drifts to finite conductance. Generally, we found the stability of the QPC to improve with time, with repeated back and forth sweeps in a limited voltage range, and with the application of moderate B-field.

%
\end{document}